\documentclass[journal]{IEEEtran}

\ifCLASSINFOpdf
\else
\fi
\usepackage[T1]{fontenc}
\usepackage[latin9]{inputenc}
\usepackage{float}
\usepackage{calc}
\usepackage{amsmath}
\usepackage{graphicx}
\usepackage{subcaption}
\usepackage{mathtools}
\usepackage[export]{adjustbox}
\usepackage{listings}
\usepackage{amssymb}
\usepackage{flexisym}
\usepackage{xcolor}
\usepackage{wrapfig}
\usepackage{textcase}
\usepackage{soul}
\usepackage{comment}
\usepackage{multirow}

 \graphicspath{{./imgs/}}
 \usepackage[ruled,vlined]{algorithm2e}
\usepackage{tabularx,booktabs}
\usepackage[numbers,sort&compress]{natbib}
\usepackage{tikz}
\usetikzlibrary{positioning}

 \newcommand{\mb}{\color{black}}
 \newcommand{\mr}{\color{red}}

\begin{document}
\title{Model-based Reconstruction for Multi-Frequency Collimated Beam Ultrasound Systems}

\author{Abdulrahman M. Alanazi$^{1,4}$, Singanallur Venkatakrishnan$^{2}$,
        Hector Santos-Villalobos$^{3}$, Gregery T. Buzzard$^{1}$,
        and~Charles Bouman$^{1}$\\
$^{1}$Purdue University-Main Campus, West Lafayette, IN 47907.\\
$^{2}$Oak Ridge National Laboratory, One Bethel Valley Road, Oak Ridge, TN 37831.\thanks{This manuscript has been supported by UT-Battelle, LLC under Contract No. DE-AC05-00OR22725 with the U.S. Department of Energy. G. Buzzard was partially supported by NSF CCF-1763896, and C. Bouman was partially supported by the Showalter Trust. The United States Government retains and the publisher, by accepting the article for publication, acknowledges that the United States Government retains a non- exclusive, paid-up, irrevocable, world-wide license to publish or reproduce the published form of this manuscript, or allow others to do so, for United States Government purposes. The Department of Energy will provide public access to these results of federally sponsored research in accordance with the DOE Public Access Plan (http://energy.gov/downloads/doe-public-access-plan).}\\
$^{3}$Amazon Prime Video, 410 Terry Ave N, Seattle 98109, WA. \\
$^{4}$King Saud University (KSU), Riyadh, Saudi Arabia.}

\maketitle

\begin{abstract}
Collimated beam ultrasound systems are a technology for imaging inside multi-layered structures such as geothermal wells.
These systems work by using a collimated narrow-band ultrasound transmitter that can penetrate through multiple layers of heterogeneous material.
A series of measurements can then be made at multiple transmit frequencies.
However, commonly used reconstruction algorithms such as Synthetic Aperture Focusing Technique (SAFT) tend to produce poor quality reconstructions for these systems both because they do not model collimated beam systems and  they do not jointly reconstruct the multiple frequencies.

In this paper, we propose a multi-frequency ultrasound model-based iterative reconstruction (UMBIR) algorithm designed for multi-frequency collimated beam ultrasound systems. The combined system targets reflective imaging of heterogeneous, multi-layered structures.
For each transmitted frequency band, we introduce a physics-based forward model to accurately account for the propagation of the collimated narrow-band ultrasonic beam through the multi-layered media.
We then show how the joint multi-frequency UMBIR reconstruction can be computed by modeling the direct arrival signals, detector noise, and incorporating a spatially varying image prior.
Results using both simulated and experimental data indicate that multi-frequency UMBIR reconstruction yields much higher reconstruction quality than either single frequency UMBIR or SAFT.
\end{abstract}


\IEEEpeerreviewmaketitle

\section{Introduction}

Non-destructive evaluation (NDE) of multi-layered structures that can be accessed from only a single side is important in many applications.
For example, this imaging scenario occurs when monitoring the structural integrity of oil and geothermal wells that lie behind layers of fluid and steel casing.
While ultrasound imaging is widely used in NDE applications, multi-layered structures present a challenge for ultrasounds systems because of the complex propagation and reverberation of the signal through the material.

\begin{figure}[!htb]
\begin{center}
\includegraphics[width=0.45\textwidth]{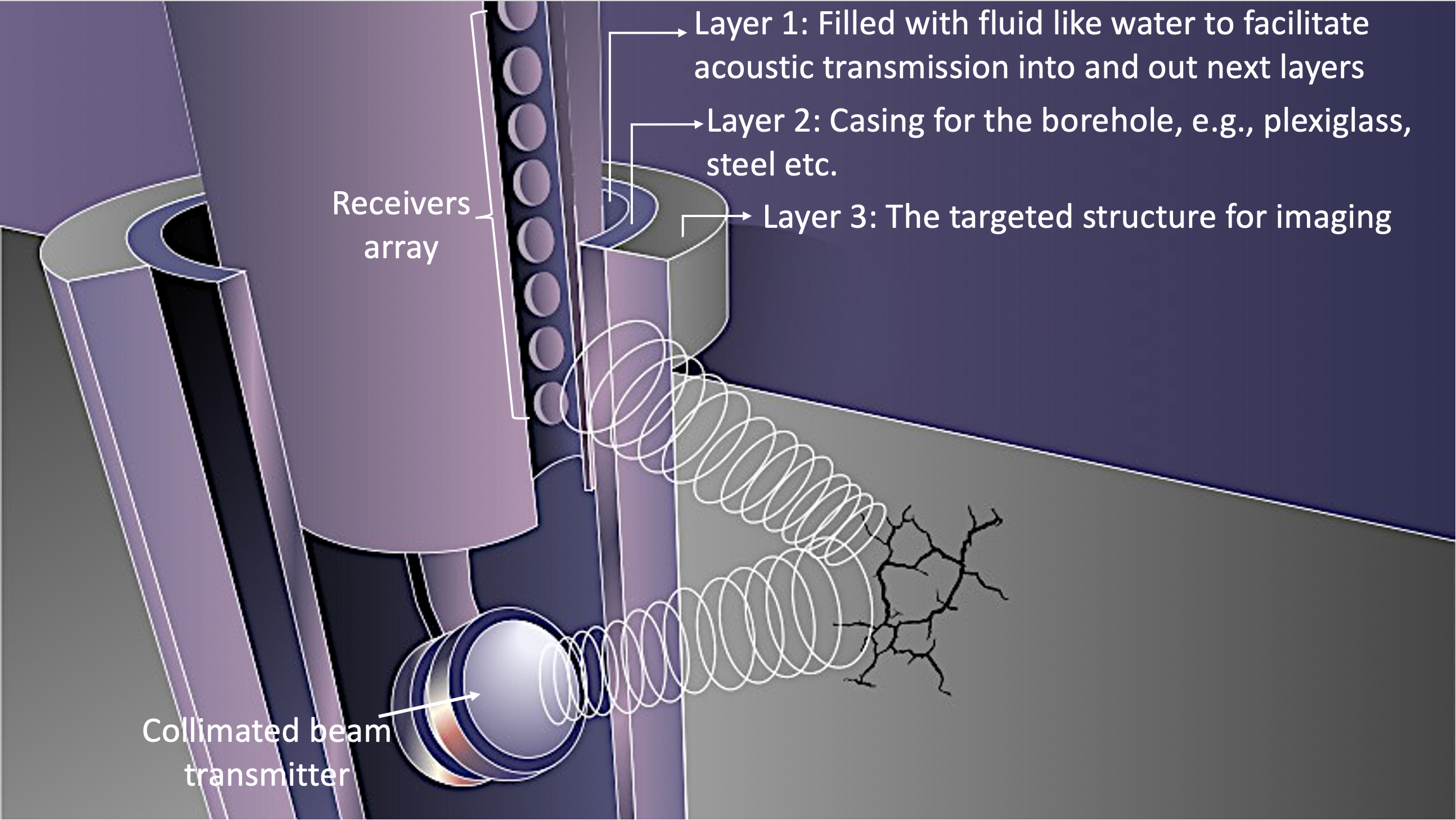}
\flushleft\caption{
Illustration of a collimated beam ultrasound system used to inspect structures that are behind multiple layers. 
The transmitter is designed to send a collimated narrow-band ultrasound signal that can penetrate a number of layers of heterogeneous material; including transitions between liquid and solid interfaces. 
Such systems are useful in applications such as sub-surface oil/geothermal wells where the goal is to image structures behind a layer of water and steel casing. 
}
\label{fig:MultiLayer_strucut_example}
\end{center}
\end{figure}

Figure~\ref{fig:MultiLayer_strucut_example} illustrates an example of a collimated beam ultrasound system \cite{chillara2017low,chillara2018radial,chillara2019collimated,chillara2020ultrasonic} that is designed to image through multi-layered structures.
The system consists of a narrow-band collimated beam transmitter combined with an array of receivers.
The system can penetrate through heterogeneous layers because the transmitter is collimated and the center frequency is below 100 kHz. 
However, since the systems are narrow-band, they typically are operated at a few different center frequencies, and then the data from each measurement frequency is processed separately to image the structure.

The most popular methods to reconstruct data from ultrasound systems use a  delay-and-sum (DAS) approach because of their low computational complexity. 
One such approach is the synthetic aperture focusing technique (SAFT), which produces acceptable ultrasound images for simple objects \cite{prine1972synthetic}. 
SAFT has been applied to single-layer \cite{stepinski2007implementation,hoegh2015extended} and multi-layered structures \cite{skjelvareid2011synthetic,lin2018ultrasonic} but not to collimated beam systems. 
Multi-layer SAFT combines DAS with techniques such as ray-tracing \cite{cerveny2005seismic,margrave2019numerical,shlivinski2007defect} and root-mean-square velocity \cite{gazdag1978wave,skjelvareid2010ultrasound} to compute the travel time in multi-layered media. 
However, SAFT and its variations rely on a simple
model that often leads to artifacts such as multiple reflections and blur. 
Methods to counteract these effects include \cite{ozkan2017inverse} and
\cite{tuysuzoglu2012sparsity}, which use a linear forward model for single-layer structures
and a carefully constructed sparse deconvolution approach.
Finally, the conventional practice when processing data obtained from multiple transmit frequency bands is to obtain the reconstruction for each band separately and then visualize the results in order to identify structures of interest. 

More physically realistic inversion methods from seismology include least-squares reverse time migration (LSRTM) \cite{liu2016least,zhang2015stable,chen2018full} and full wave inversion (FWI) \cite{zeng2017guide}. 
LSRTM and FWI are iterative methods that seek the best least-squares fit between observed and reconstructed data; a reflectivity image for LSRTM and a velocity image for FWI. 
These methods have the capability to image complex structures but rely on an iteration using a non-linear forward model, making them computationally expensive and impractical for many imaging applications.
Finally, these methods are typically applied separately to data from each frequency band which does not allow the end user to obtain a single image corresponding to the structure to be imaged. 

In order to reduce the reconstruction artifacts of SAFT while maintaining computational efficiency, model-based iterative reconstruction (MBIR) can be used with a linear propagation model. 
These methods combine a forward model for the ultrasound measurement system with a prior-model/regularizer for the unknown structure and cast the reconstruction as a maximum a-posteriori estimation problem. 
In \cite{ozkan2017inverse,wu2015model}, the forward model is designed to handle plane-wave imaging. 
Recently, the MBIR approach of \cite{almansouri2018model} used a propagation model of the ultrasound through the media and combined all the data from the source-detector pairs to jointly reconstruct a fully 3D image. 
However, current MBIR approaches have limitations when imaging through heterogeneous materials. 
In a single-layer structure, the time delay can be easily computed using Snell's law \cite{almansouri2018model,almansouri2018anisotropic}. 
However, in structures containing two layers, the time delay no longer has a closed form that can be easily computed using Snell's law \cite{weston2012time}.
When the number of layers exceeds two, some techniques such as ray-tracing or marching methods \cite{moser1991shortest,sethian1996fast,brath2017phased} can be used to approximate the computations but these are computationally complex.
In summary, no existing regularized MBIR method accounts
for multi-frequency collimated beam systems used to image structures which are behind several layers.

In this paper, we propose a MBIR algorithm for multi-frequency collimated beam ultrasound systems  in order to image structures that are behind multiple layers of materials. 
Our method, which we refer to as multi-frequency ultrasound model-based iterative reconstruction (UMBIR), can be used to accurately reconstruct heterogeneous structures behind multiple layers of material excited at multiple frequencies. 
UMBIR does this by combining a novel physics-based forward model that models multi-layered heterogeneous structures while also coherently integrating data from of multiple frequencies to form a single reconstructed cross section. 
In order to estimate the travel time of the ultrasound signals through multiple media as a part of the forward model, we introduce a efficient binary search-based method. 
Finally, the maximum a posteriori (MAP) estimate is then computed using iterative coordinate descent (ICD) optimization with a spatially varying prior similar to \cite{almansouri2018anisotropic}.  
We note that our work builds on preliminary ideas developed in the context of single band measurements which was presented as a conference article \cite{alanazi2022model}.
Using experimental and simulated data, we demonstrate that the proposed multi-layer and multi-frequency UMBIR approach yields more accurate reconstructions with higher spatial resolution and reduced artifacts when compared to both single frequency UMBIR and SAFT.

\section{The UMBIR Forward Model}
\label{sec:FW}

In order to image the structure which is separated from the source by multiple layers of known materials, we design a ultrasound model-based iterative reconstruction (UMBIR) approach. 
Assuming a linear system for simplicity, we seek to reconstruct an image $x$ using a measurement model of the form 
\begin{equation}
    \label{eq:sysmodel}
    y = Ax + Dg + w, 
\end{equation}
where $y\in \mathbb{R}^{MK \times 1}$ is a vector of measurements from $K$ receivers at $M$ timepoints, $A \in \mathbb{R}^{MK\times N}$ is the system matrix, $x\in \mathbb{R}^{N \times 1}$ is the vectorized version of the desired image with $N$ total voxels, $D\in \mathbb{R}^{MK \times K}$ is a matrix whose columns form a basis for the possible direct arrival signals, $g \in \mathbb{R}^{K \times 1}$ is a scaling coefficient vector for $D$, and $w$ is a Gaussian random vector with distribution $N(0, \sigma^2 I)$. 
From \eqref{eq:sysmodel}, we will be able to formulate the reconstruction as the maximum {\em a posteriori} estimate of $x$ and $g$ given $y$.
However, to do this, we will first need to introduce an acoustic model of propagation through the multi-layered material that we can use to compute the matrices $A$ and $D$.

\begin{figure}[htb]
\begin{center}
\includegraphics[width=0.45\textwidth]{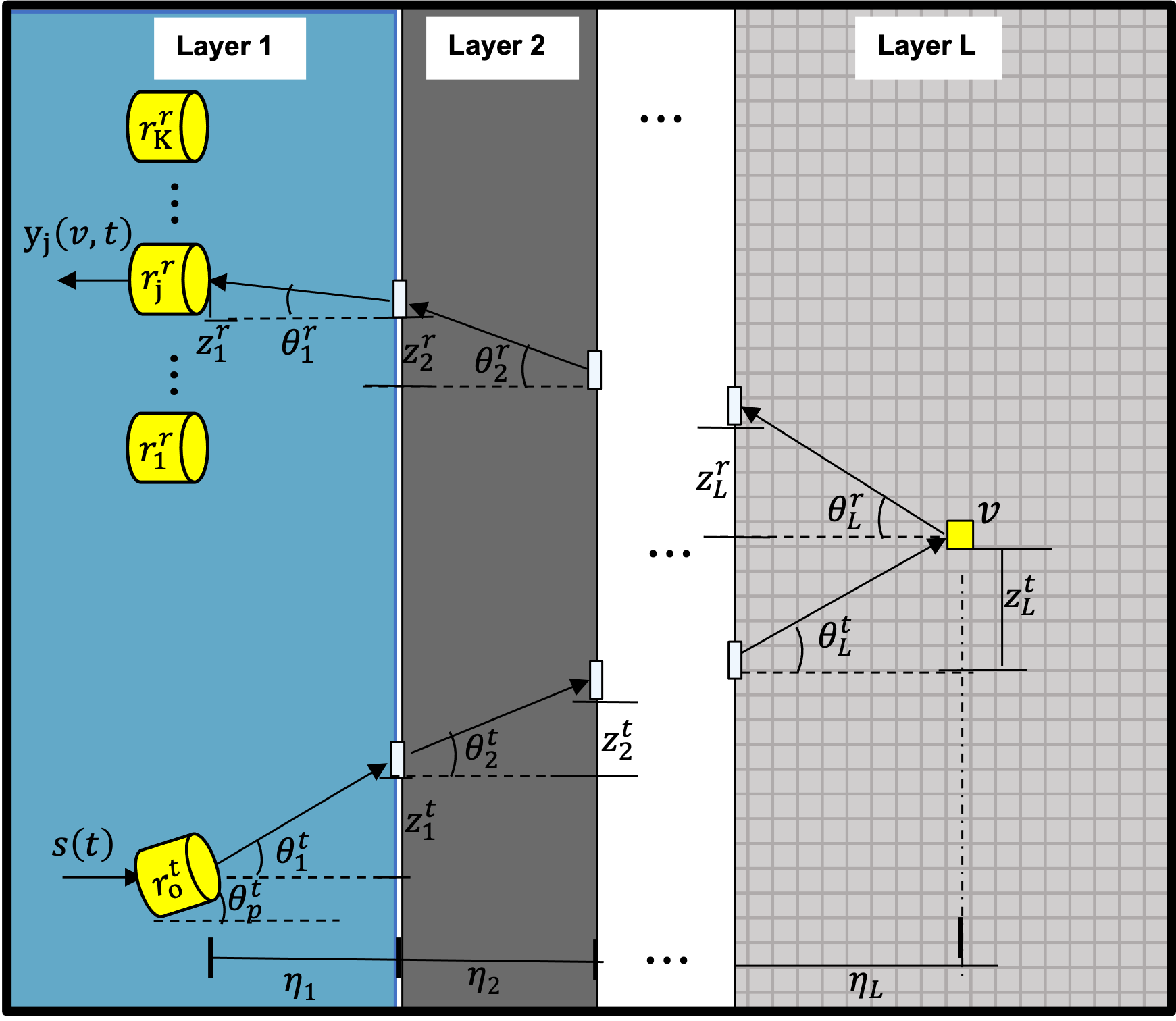}
    \flushleft\caption{Illustration of method used to compute time delays of ultrasound signals propagating through a multi-layered medium. 
    $s(t)$ is the input signal and $y_j(v,t)$ is the received signal at the $j^{th}$ microphone due to reflection from voxel $v$.}
    \label{fig:MultiLay}
    \end{center}
\end{figure}

\subsection{Multi-Layer Acoustic Propagation Model}
\label{sec:multilayer-acoustic-model}

In this section, we introduce a model of the multi-layer acoustic propagation based on an extension of the single-layer model used in~\cite{almansouri2018model}.
Figure~\ref{fig:MultiLay} illustrates the problem.
An acoustic signal is transmitted from location $r^t_o$, reflected from the voxel, $v$, and then received by one of $K$ possible microphones at location $r^r_j$.
As the signal propagates outward, it passes through $L$ different materials each with its own acoustic velocity, $c_l$, in $\frac{m}{s}$, and attenuation coefficient, $\alpha_l$, in $\frac{s}{m}$.
Let $T^t_\ell (v)$, and $T^r_{\ell,j}(v)$, denote that outgoing and returning propagation time of the beam through each layer of the medium. Notice that both times will be functions of the particular voxel, $v$, and the return time will also be a function of the particular microphone, $j$.
Then we model the frequency domain transfer function as
\begin{equation}
    \label{eq:TransFerfunction1}
    G_j (v,f) = \lambda (v) \prod_{\ell=1}^{L} \mr  \mb e^{-( c_\ell \alpha_\ell |f|+2 \mathfrak{j} \pi f ) \left[ T^t_\ell (v) + T^r_{\ell,j}(v)\right] } \ , 
\end{equation}
where $\mathfrak{j}^2 = -1$ and 
\begin{equation}
    \label{eq:TransCoeff}
    \lambda (v) = \phi_j (v) \left( \prod_{\ell=2}^{L} \frac{2 \zeta_\ell}{\zeta_{\ell-1}+\zeta_\ell} \right) \left( \prod_{\ell=1}^{L-1}   \frac{2 \zeta_\ell}{\zeta_{\ell}+\zeta_{\ell+1}} \right) \ ,
\end{equation}
where $\zeta_\ell$ in $\frac{kg}{m^2s}$ is the acoustic impedance of the $\ell^{\text{th}}$ layer and $\phi_j (v)$ is the scalar correction due to beam collimation discussed in Section~\ref{sec:beam-collimation}.
 Note that $\lambda (v)$ in \eqref{eq:TransFerfunction1} models the effects on the amplitude of the received signal by the beam collimation and impedance mismatch between the layers, which are the formulas defined between parentheses in \eqref{eq:TransCoeff}. 
We can further simplify the expression, but defining two quantities
\begin{align}
\gamma_j (v) &= \sum_{l=1}^L c_{\ell} \alpha_{\ell} \left[ T^t_\ell (v) + T^r_{\ell,j}(v)\right] \\
T_j (v) &= \sum_{l=1}^L \left[ T^t_\ell (v) + T^r_{\ell,j}(v)\right] \ .
\end{align}
In this case, the frequency domain transfer function has the simpler form of
\begin{equation}
    \label{eq:SimplifiedTransFerfunction1}
    G_j (v,f) = \lambda (v) \, e^{-( \gamma_j (v)\, |f|+2 \mathfrak{j} \pi f \, T_j (v) ) } \ .
\end{equation}
In this form, it is clear that $T_j (v)$ represents the round-trip group delay to voxel $v$, and $\gamma_j (v)$ represents the signal dispersion.
From this, the Fourier transform of the received signal given by 
\begin{equation}
    \label{eq:receivedSignal_freq}
    Y_{j}(v,f) = x(v) \lambda (v) S(f) e^{-( \gamma_j (v)\, |f|+2 \mathfrak{j} \pi f \, T_j (v) ) } \ ,
\end{equation}
where $S(f)$ is the Fourier transform of the transmitted signal, and $x(v)$ is the reflection coefficient for the voxel $v$.
Then in the time-domain the received signal is given by
\begin{equation}
    \label{eq:receivedSignal_time}
    y_{j}(v,t) = x(v)\, \lambda (v) \, h(\gamma_{j} (v),t-T_{j}(v)),  
\end{equation}
where 
\begin{equation}
    \label{eq:h}
    h(\gamma ,t) = \mathcal{F}^{-1}\left\{ S(f) e^{-\gamma |f|} \right\}  
\end{equation}
and $\mathcal{F}^{-1}$ is the inverse Fourier transform. 

Thus, the output at each time $t$ and each receiver $j$ can be expressed as an inner product between the input $x(v)$ and a row of the forward model's system matrix, $A$.
This is expressed in the following formula.
\begin{align}
    \label{eq:y_apprx}
    y_{j}(t) &= \sum_v \left[ h(\gamma_{j} (v),t-T_{j} (v)) \, \lambda (v) \right] x(v) 
\end{align}

So from this we see that, for the multi-layer case, we will need to first compute, $T_{j} (v)$, the acoustic travel time of the signal through each layer for each voxel in order to obtain a expression for the received signal. 
This is the subject of the next section.

\subsection{Time Delay Computation} 
\label{sec:TimeDelay}

In single-layer structures, the computation of the time delays at the image voxels is straightforward. 
However, in multilayered structures, the acoustic speed varies with depth, which causes reflections and refractions that result in a complex wave path. 
In order to compute the received signal of~\eqref{eq:y_apprx}, we need to know the time delays for each layer, $T^t_{\ell}(v)$ and $T^r_{\ell , j}(v)$, which will be dependent on the path of the acoustic signal through the multiple layers of the material.
Figure~\ref{fig:MultiLay} illustrates the acoustic path as it passes through the layers of the medium. Because each layer has different acoustic velocity, $c_\ell$, the signal will be refracted as it passes between layers. 
Let $\theta^t_{\ell}$ represent the angle of outbound propagation through the $\ell^{th}$ material, and let $\theta^r_{\ell,j}$ represent the angle of return propagation, as in Figure~\ref{fig:MultiLay}. Then by Snell's law, we know that 
\begin{equation}
    \label{eq:thetas_out}
    \theta^t_{\ell } = \text{sin}^{-1}\left( \text{sin}(\theta^t_{\ell -1}) \frac{c_\ell }{c_{\ell -1}} \right) \ ,
\end{equation}
and the return angles are given by
\begin{equation}
    \label{eq:thetas_return}
    \theta^r_{\ell -1 } = \text{sin}^{-1}\left( \text{sin}(\theta^r_{\ell}) \frac{c_\ell }{c_{\ell + 1}} \right) \ .
\end{equation}
Consequently, if we know $\theta^t_{1 }$ and $\theta^r_{L }$, then we can use the recursions of \eqref{eq:thetas_out} and~\eqref{eq:thetas_return} to compute the remaining angles. We denote these two functions that represent the result of this calculation as
\begin{eqnarray*}
\theta^t_{\ell } &=& f^t_\ell \left[ \theta^t_1 \right] \\
\theta^r_{\ell } &=& f^r_\ell \left[ \theta^r_L \right] \ .
\end{eqnarray*}
Let $\eta_\ell $ denote the thickness of the $\ell^{th}$ layer, then we can express the vertical distance that the outbound acoustic signal travels as 
\begin{align}
\label{eq:height-of-v}
Z^t \! \left[ \theta^t_1 \right] = 
\sum_{\ell=1}^{L} \eta_{\ell } \text{tan} \left( f^t_{\ell } \left[ \theta^t_1 \right] \right) \ , 
\end{align}
and the vertical distance that the returning signal travels as
\begin{align}
\label{eq:height-of-v}
Z^r \! \left[ \theta^r_L \right] = 
\sum_{\ell=1}^{L} \eta_{\ell } \text{tan} \left( f^r_{\ell } \left[ \theta^r_L \right] \right) \ . 
\end{align}

\begin{figure}[htb]
\centering
\begin{tabular}{ccccc}
\tabularnewline
\includegraphics[width=0.235\textwidth]{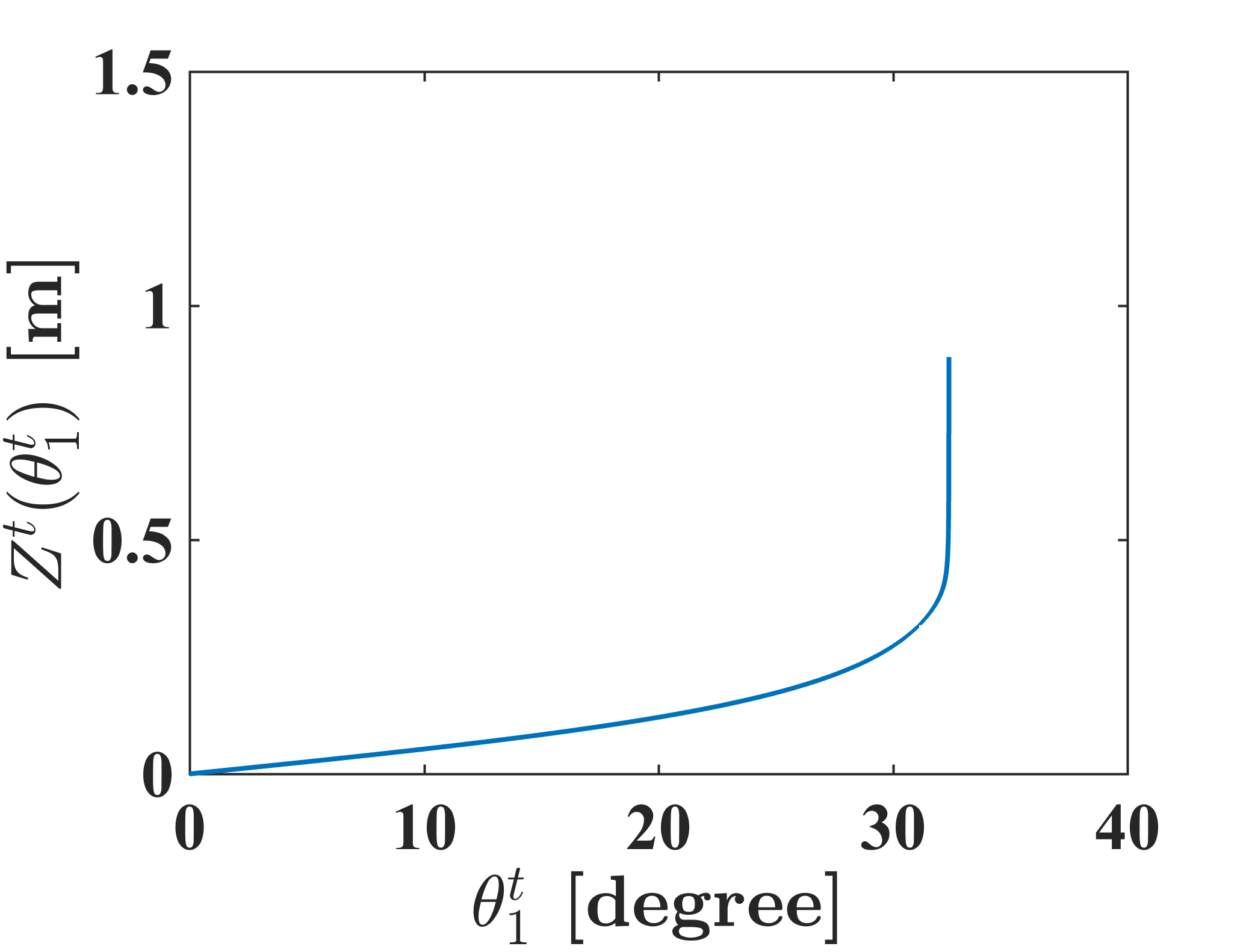} 
\includegraphics[width=0.235\textwidth]{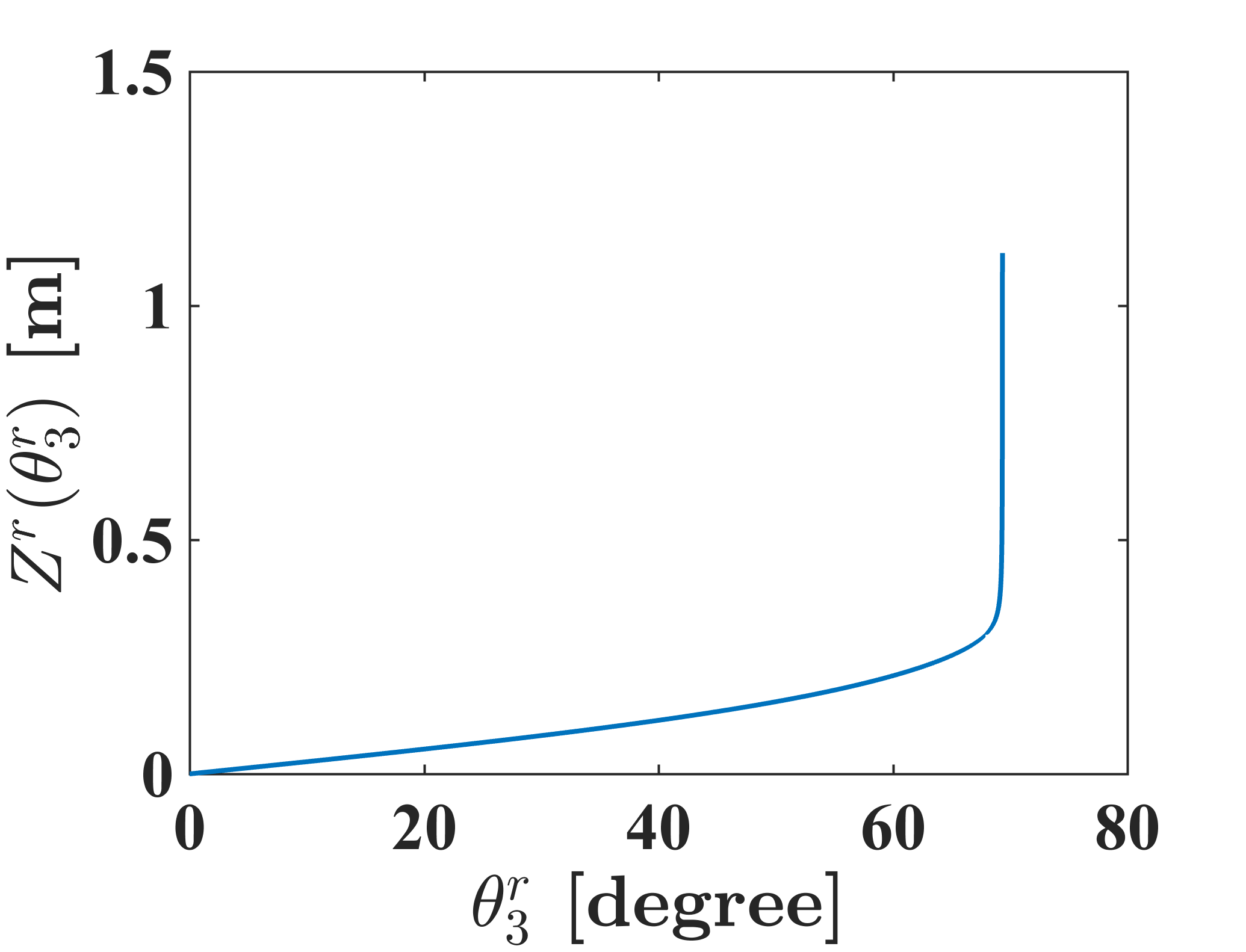} 
 \tabularnewline
(a)  \quad \quad \quad \quad \quad \quad  \quad \quad \quad \quad \quad (b)
 \end{tabular}
 \caption{(a) The vertical distance traveled by the outgoing acoustic signal, $Z^t(\theta^t_1)$, as a function of $\theta^t_1$. (b) The vertical distance traveled by the returning signal, $Z^r(\theta^r_3)$, as a function of $\theta^r_3$.
Notice that both functions are monotone increasing.
}
\label{fig:h_theta}
\end{figure}

As an example, Figure \ref{fig:h_theta} shows the relationship between $Z^t(\theta^t_1)$ versus $\theta^t_1$ and $Z^r(\theta^r_3)$ versus $\theta^r_3$ in three layer media with parameters as shown in Table~\ref{table:Multi-Layer-Media-Params}.
We note that it is clear from this example, that $Z^t(\theta^t_1)$ is a monotone increasing function of $\theta^t_1$. 

\begin{table}[htb]
\caption{The parameters used to show the relationship between the vertical distance and propagation angle in Figure~\ref{fig:h_theta}.} 
\label{table:Multi-Layer-Media-Params}
\centering
\begin{tabular}{l|l|l|l|}
\cline{2-4}
                                                                             & \textbf{Layer 1} & \textbf{Layer 2} & \textbf{Layer 3} \\ \hline
\multicolumn{1}{|l|}{\textbf{Thickness ($\eta_\ell$) in m}}                  & 0.073            & 0.006            & 0.12             \\ \hline
\multicolumn{1}{|l|}{\textbf{Acoustic velocity ($c_\ell$) in $\frac{m}{s}$}} & 1500             & 2800             & 2620             \\ \hline
\end{tabular}
\end{table}

Next, for each voxel, $v$, we must solve for the unknown angles, $\theta^t_{1 }(v)$, the departure angle of the outbound acoustic signal, and $\theta^r_{L }(v)$, the arrival angle of the returning acoustic signal.
We can do this by solving the following two equations.
\begin{eqnarray*}
Z_v - Z_o &=& Z^t \left[ \theta^t_1 (v) \right] \\
Z^r_j - Z_v &=& Z^r \left[ \theta^r_{L,j} (v) \right] \ ,
\end{eqnarray*}
where $Z_v - Z_o$ is the vertical distance between the transmitter, $r^t_o$, and the voxel, $v$; and $Z^r_j - Z_v$ is the vertical distance between the microphone, $r^r_j$, and the voxel, $v$.
Since both functions are monotone increasing functions of their arguments, these equations can be easily solved using half interval search as described in \cite{bouman2022foundations}.

Once the values of $\theta^t_{1 }$ and $\theta^r_{L }$ are determined, then the values of the time delay can be computed using the following two equations.
\begin{eqnarray}
\label{eq:TD-outbound}
T^t_{\ell } (v) &=& \eta_\ell \frac{ \sqrt{ 1 + \tan^2 \left( f^t_\ell [ \theta^t_1 (v) ] \right) } }{c_\ell} \\[15pt]
\label{eq:TD-return}
T^r_{\ell, j} (v) &=& \eta_\ell \frac{ \sqrt{ 1 + \tan^2 \left( f^r_\ell [ \theta^r_{L,j} (v) ] \right) } }{c_\ell} \ .
\end{eqnarray}
This provides all the values that are needed to compute the solution to \eqref{eq:SimplifiedTransFerfunction1}.

\subsection{Collimated Beam Modeling}
\label{sec:beam-collimation}

In this section, we develop a model for the collimated beam generated by the acoustic source transducer used in our experiments. 
In order to accurately model the effects of beam collimation, we use an apodization function inspired by work of \cite{hoegh2015extended} as later adapted by \cite{almansouri2018model}.  
Adapting this approach to better model the collimated beam and defining the angle at the receiver as $\theta^r_{1,j} = f_1^r[\theta^r_{L,j}(v)]$ our apodization function is given by
\begin{equation}
\label{eq:apodFunc}
\phi_j (v) = 
\text{cos}^\beta \left( \theta^t_1 (v) - \theta^t_p \right) 
\cos^2 \left( \theta_{1,j}^r(v) \right)
\ ,
\end{equation}
where $\beta$ is a parameter that controls the beam apodization, and $\theta^t_p$ is the pointing angle of the transmitter as shown in Figure~\ref{fig:MultiLay}. 
The reason behind modeling the pointing angle of the transmitter is because in practice, the transmitter is often tilted upward with respect to the axial direction of the microphones array to increase the ultrasonic illumination range and ensure reflection of the transmitted signal to the receivers \cite{chen2018full}.

\subsection{Single Frequency System Matrix Construction}
\label{sec:SystemMatrixConstruction}

In this section, we describe how the the matrices, $A$ and $D$, are constructed from the multi-layer acoustic model for a single frequency as described in Section~\ref{sec:FW}. 
In the following section, we will describe how these system matrices are combined to form the full system matrix used in the multi-frequency case.

In order to reduce computation, we window our model of the received signal in time by replacing $h$ of~\eqref{eq:receivedSignal_time} with
\begin{equation}
    \label{eq:h_apprx}
    \tilde{h} (\gamma_{j} (v) ,t) = h(\gamma_{j} (v),t) \, \text{rect}\left( \frac{t}{t_0} - \frac{1}{2} \right),  
\end{equation}
where $t_0$ is a constant based on the assumption that $h(\gamma_{j}(v),t)\approx 0$ for $t > t_0$ and $\text{rect}(u) = 1$ for $|u| < \frac{1}{2}$ and is $0$ otherwise.
By populating $A$ and $D$ with the windowed function, $\tilde{h}$, we ensure that the matrices are sparse so that computation and memory usage are reduced.

Equation~\eqref{eq:y_apprx} can then be used to populate the entries of the system matrix $A$.
For each voxel, $v_i$, and receiver location, $r_j^r$, the following partial column vector is formed
\begin{align}
\label{eq:partial_col_A}
a^{i,j}  = 
\left[ 
\begin{array}{c}
\tilde{h} (\gamma_{j} ( v_i ) ,t_0 )  \\
\vdots \\
\tilde{h} (\gamma_{j} ( v_i ) ,t_{M-1} ) 
\end{array}
\right]
\end{align}
where $t_m= m \Delta + T_o$ and $\Delta$ is the time sampling period.
A full column vector of the system matrix, $A$, is then formed by concatenating the partial columns for each receiver.
\begin{align}
\label{eq:full_col_A}
A_{*,i}  = 
\left[ 
\begin{array}{c}
a^{i,1}  \\
\vdots \\
a^{i,K} 
\end{array}
\right]
\end{align}
And then the full system matrix is formed by concatenating the columns.
\begin{align}
\label{eq:full_A}
A  = \left[ A_{*,1}, \cdots , A_{*,N} 
\right]
\end{align}

Note that Equation~\eqref{eq:y_apprx} can also be used to populate the entries of the matrix $D$. However, the group delay and signal dispersion formulas introduced above depend on the voxel $v$; which is not the case for the direct arrival signals. Hence, we define the time delay from the transmitter, $r^t_o$, to each receiver, $r^r_j$, and the direct arrival signal dispersion, respectively, as 
$$
\tau_j = \frac{\lVert r^t_o - r^r_j \rVert}{c_d} \ ,
$$
$$
\Bar{\gamma_j} = \alpha_d c_d \tau_j \ ,
$$
where $\alpha_d$ and $c_d$ are the attenuation coefficient and acoustic velocity of the material that the transducers are embedded in. 
The matrix $D$ is formed from $K$ columns, one for each detector location $r^r_j$.
We denote the columns of $D$ as $d^j$.
Then, for the $k^{th}$ receiver, a partial column vector, $d^k\in \Re^M$, is formed by
\begin{align}
\label{eq:partial_col_D}
d^k = 
\left[ 
\begin{array}{c}
\tilde{h} (\Bar{\gamma_j} ,t_0 )  \\
\vdots \\
\tilde{h} (\Bar{\gamma_j} ,t_{M-1} ) 
\end{array}
\right] \ .
\end{align}
These partial vectors can then be concatenated to form the full matrix given by
\begin{align}
\label{eq:full_D}
D  = \left[ 
\begin{array}{cccc}
d^{1} & 0 & \cdots & 0  \\
0 & d^{2} & \cdots & 0  \\
\vdots & \vdots & \vdots & \vdots \\
0 & 0 & \cdots & d^{K}  
\end{array}
\right]
\end{align}

\subsection{Multi-Frequency UMBIR}
\label{sec:joint}

In order to do multi-frequency reconstruction, we must form a system matrix that account for measurements at all frequencies simultaneously.
Let $S$ denote the number of distinct excitation frequencies, and let $A^s$ and $D^s$ denote the associated system matrix and direct arrival signal matrix constructed using the methods described in Section~\ref{sec:SystemMatrixConstruction}, and let $y_s$ denote the measurements associated measurements.
Then we can form the full measurement vector, $y$, full system matrix, $A$, and direct arrival signal matrix, $D$, to be as follows
\begin{align}
\label{eq:full_y_multifrequency}
y = \left[ 
\begin{array}{c}
y^1 \\
\vdots \\
y^S
\end{array}
\right] 
\hspace*{10pt}
A = \left[ 
\begin{array}{c}
A^1 \\
\vdots \\
A^S
\end{array}
\right] 
\hspace*{10pt}
D = \left[ 
\begin{array}{cccc}
D^1 & \cdots & 0 \\
\vdots & \ddots & \vdots \\
0 & \cdots & D^S
\end{array}
\right]. 
\end{align}

\section{Prior Model of UMBIR}
\label{sec:PriorModel}

For the prior model, we adopt the q-generalized Gaussian Markov random field (QGGMRF) from \cite{thibault2007three,almansouri2018model}. With this design, the prior probability is
\begin{eqnarray}
\label{eq:prior}
p(x) = \frac{1}{z}\exp\left(- \sum_{\{s,r\} \in C} b_{s,r} \ \rho(x_s-x_r)\right),
\end{eqnarray}
where $z$ is a normalizing constant, $C$ is the set of pair-wise cliques, 
\begin{eqnarray}
\rho(\Delta) = \frac{|\Delta|^p}{p\sigma_{s,r}^p}\left( \frac{|\frac{\Delta}{T\sigma_{g_{s,r}}}|^{q-p}}{1+|\frac{\Delta}{T\sigma_{s,r}}|^{q-p}}\right), \label{pot}
\end{eqnarray}
\begin{eqnarray}
\sigma_{s,r} &=& \sigma_{0} \sqrt{\nu_s \nu_r}\label{SVR2},
\end{eqnarray}
\begin{eqnarray}
\nu_{s} &=& 1 + (\nu-1)*\left( \frac{d_s}{d_{max}} \right) ^a, \label{eq:SVR}
\end{eqnarray}
where $\nu>0$, $a>0$, $d_s$ is the distance from the sensor assembly to pixel $s$, and $d_{max}$ is the max of $d_s$ over all $s$. 
We use $1<p<q = 2$ to insure convexity and continuity of first and second derivatives of the prior model. 
The parameter $T$ is unit-less and controls the edge threshold. The QGGMRF parameter  $\nu$ is unit-less and can be adjusted to amplify reflections at deeper regions if needed. 
Finally, by taking the negative log of Equation (\ref{eq:prior}), the prior model penalty function is given by
\begin{eqnarray}
-\log p(x) = \sum_{\{s,r\} \in C} b_{s,r} \ \rho(x_s-x_r) + \text{constant} \ .
\label{eq:NegLogPriorTerm}
\end{eqnarray}

\section{MAP Estimation and the UMBIR Algorithm}
\label{sec:MAP}

In order to perform UMBIR reconstruction, we will need to estimate both the image, $x$, and the direct arrival signals coefficients, $g$.
Using the MAP formulation, the multi-layer UMBIR reconstruction is then given by
\begin{align}
\label{x_MAP}
& \left( \hat{x}, \hat{g} \right) \nonumber \\
&= \arg \min_{ (x, g) } \left\{ -\log p(y|x,g ) - \log p(x) -\log p_g (g) \right\} \ .
\end{align}
We will use an improper prior distribution for $g$ of the form 
$$
-\log p_g (g)= \text{constant} \ ;
$$
the prior term for $x$ is given by~\eqref{eq:NegLogPriorTerm} above;
and the forward model term is given by
\begin{equation}
\label{eq:FW}
-\log p(y|x, g) = \frac{1}{2\sigma^2} \left\|  y-Ax-Dg\right\|_{2}^{2} + \text{constant} \ ,
\end{equation}
where the system matrix, $A$, and the direct arrival matrix, $D$, are constructed as described in Section~\ref{sec:FW} above.

Putting this together, results in the 
\begin{align}
(\hat{x},\hat{g}) = \arg \min_{ (x, g) }
&\left\{ \rule{0pt}{20pt} \frac{1}{2\sigma^2} \left\|  y-Ax-Dg\right\|^2 \right. \nonumber \\
& \left. + \sum_{\{s,r\} \in C} b_{s,r} \ \rho(x_s-x_r) \right\}.
\label{eq:final_map}
\end{align}
In order to compute the MAP estimate \eqref{eq:final_map}, we use Iterative Coordinate Descent (ICD) algorithm with the majorization technique for the prior model as described in \cite{bouman2022foundations}.

\begin{figure*}[htb]
 \centering
\begin{tabular}{ccccc}
\tabularnewline
\includegraphics[width=0.2\textwidth]{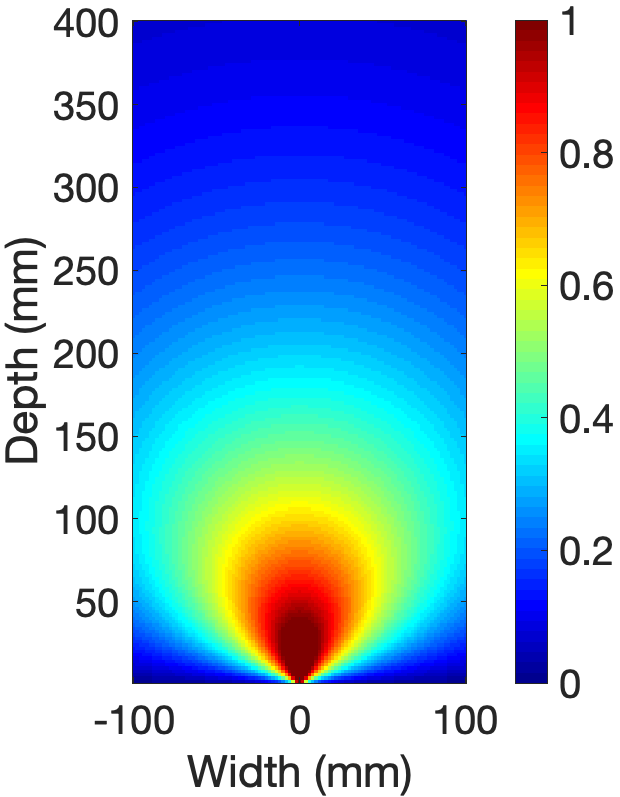} &
\includegraphics[width=0.2\textwidth]{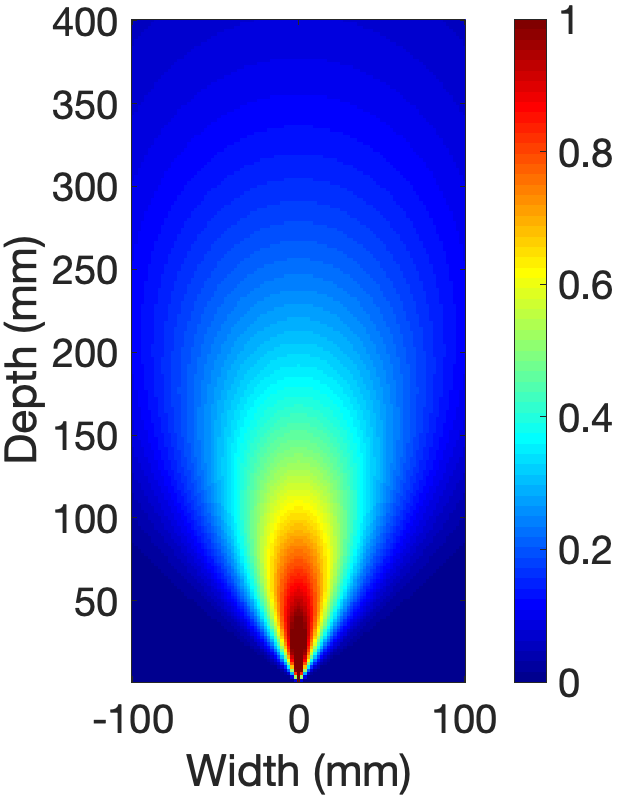} &
\includegraphics[width=0.2\textwidth]{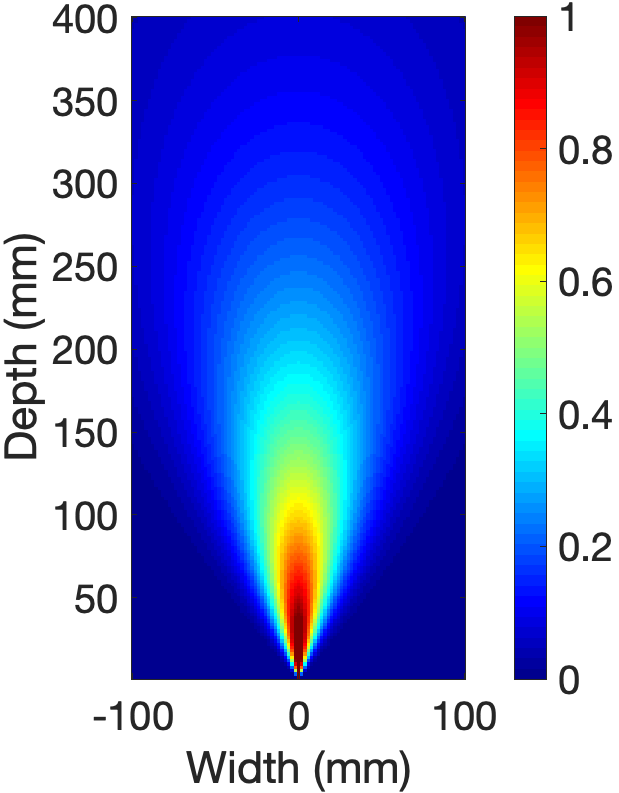} &
 \includegraphics[width=0.2\textwidth]{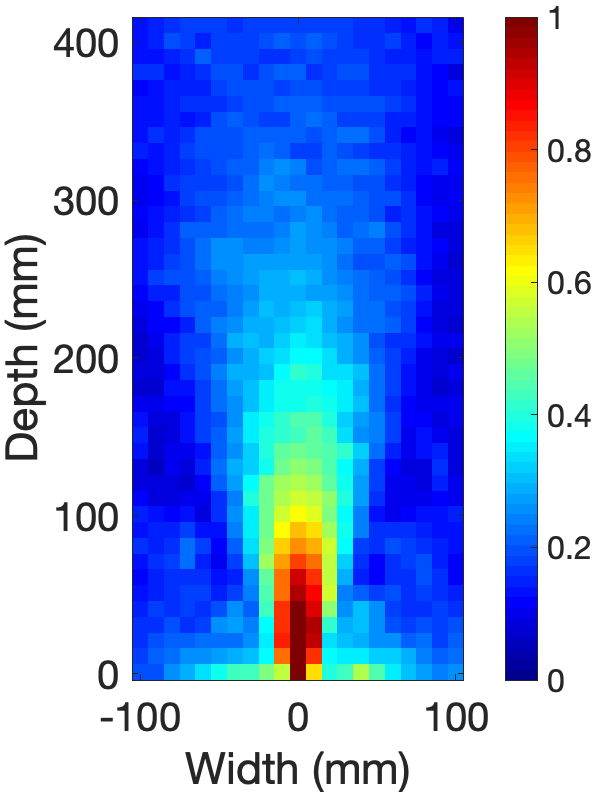} 
 \tabularnewline
(a) $\beta=1$ & (b) $\beta=4$ & (c) $\beta=8$ & (d) Measured 
 \tabularnewline
\end{tabular}
\caption{
Comparison of experimentally measured and simulated apodization functions.
A simulated beam profile, $\phi(v)^{(\beta)}e^{-\alpha_0 r(v)}$, using the apodization function of \eqref{eq:apodFunc} for
(a) $\beta=1$, (b) $\beta=4$, and (c) $\beta=8$.
(d) Experimentally measured beam profile for a collimated source.
Notice that the value of $\beta=8$ provides a reasonably accurate approximation to the true beam profile.
}
\label{fig:beams}
\end{figure*}
\begin{figure*}[!htb]
\centering
\begin{tabular}{cc}
\tabularnewline
\includegraphics[height=6cm]{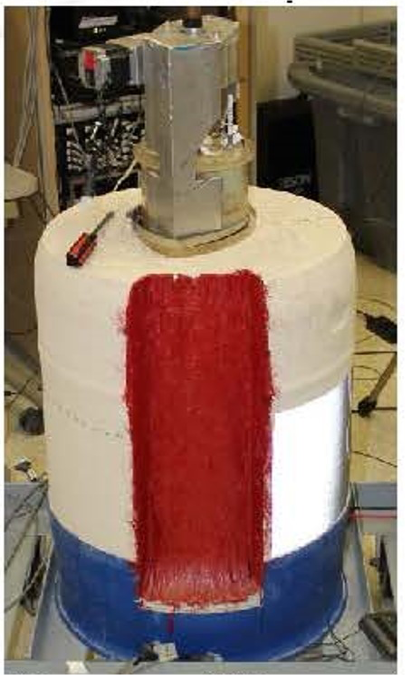} &
\includegraphics[height=6cm]{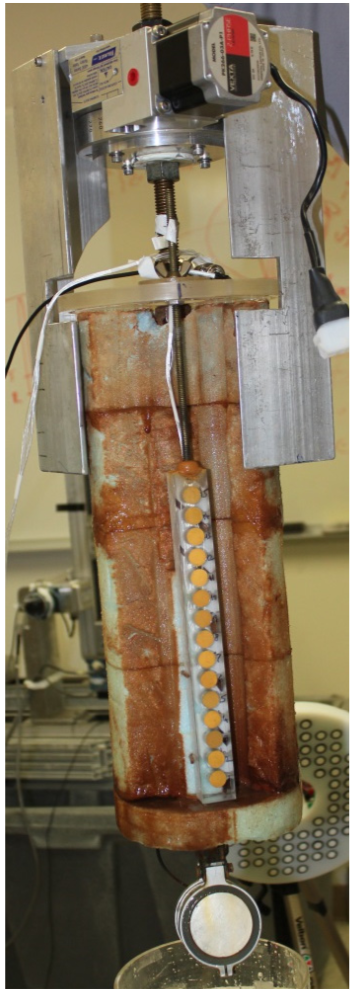}\\
  (a) & (b) \\[15pt]
\includegraphics[height=9cm]{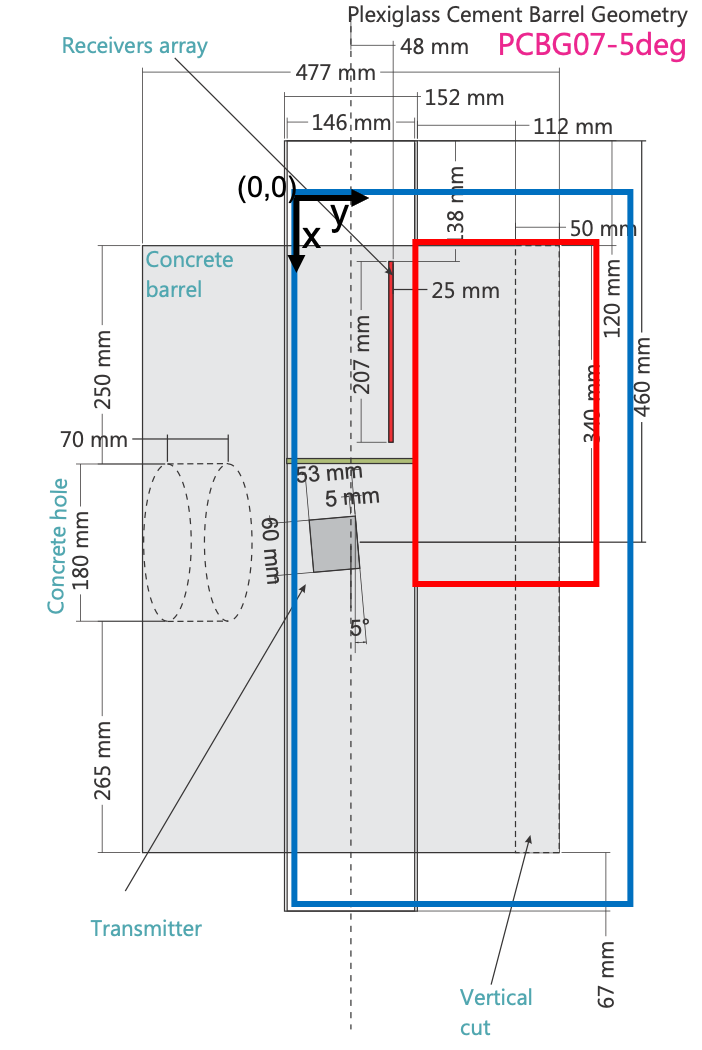} &
\includegraphics[height=8cm]{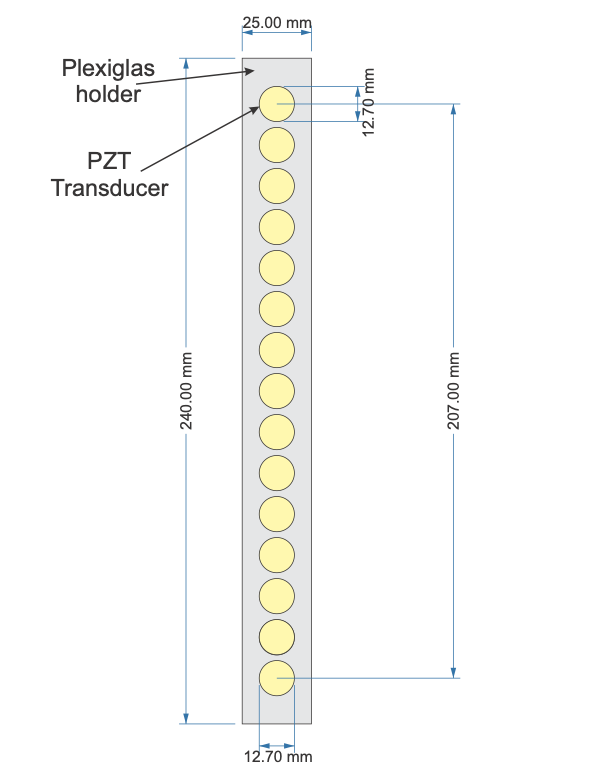} \\[-5pt]
  (c) & (d)
\end{tabular}
\caption{Illustration of experimental set up for concrete cylinder experiment.
(a) Picture of concrete cylinder that contains a central bore with the acoustic imaging senor. Notice that a 50 mm notch in the concrete is painted red.
(b) Picture of acoustic imaging sensor removed from cylinder.
(c) Diagram of a cross section of the concrete cylinder with detailed specification of distances.
The blue rectangle indicates the region used to generate synthetic data to evaluate the UMBIR algorithm. 
The red region indicates the cross-section that is to be reconstructed. 
(d) Diagram of the receiver for the acoustic imaging sensor with detailed specification of sensor positions.
}
\label{fig:Concrete-Cylinder-Experiment}
\end{figure*}

\section{Experimental Results}

\label{sec:expr_results}

In this section, we present results using both synthetic and measured data sets.
The synthetic data was generated using the K-Wave simulation package \cite{treeby2010k}. 
Both our synthetic and measured data experiments are designed to evaluate the performance of the well-bore integrity inspection system designed at the Los Alamos National Laboratory (LANL) and shown in Figure~\ref{fig:Concrete-Cylinder-Experiment} and Figure~\ref{fig:LANL_II_obj}.
This system uses a collimated acoustic transmitter along with an array of 15 receiving transducers.
More detailed about the designed transducer can be found in~\cite{pantea2019collimated}.

\subsection{Methods}

In order to validate our method, we performed two different experimental studies: The concrete cylinder (CC) experiment and the granite block (GB) experiment. 
In order to evaluate our method we also generate simulated data that is similar to the CC experiment using the K-Wave software (referred to in the results as CC-KWave).
Table~\ref{table:UMBIR-Reconstruction-Params} provides all parameters used in our reconstruction experiments for the three cases.

\begin{table}[htb]
\caption{Parameter settings for concrete cylinder and granite block reconstructions.}
\label{table:UMBIR-Reconstruction-Params}
\begin{center}
\begin{tabular}{|c|c|c|c|c|}
\hline
\multicolumn{5}{|c|}{\textbf{Forward Model Parameters}} \\
\hline
\textbf{Parameter}& \textbf{CC-KWave} & \textbf{CC-Exp} & \textbf{GB-Exp}     & \textbf{Unit} \\ \hline
Num.  rows    & 140 & 140 &       110          & -             \\ \hline
Num. cols & 70 & 70 &      124           & -             \\ \hline
Recon. resolution & 3 & 3 & 3            & mm             \\ \hline
Sampling freq.& 2 & 2 &       5          & MHz             \\ \hline
FOV height        & 420& 420 &     330            & mm             \\ \hline
FOV depth         & 210 & 210 &      370           & mm             \\ \hline
$\beta$           & 8 & 8 & 8     & -   \\ \hline
$\alpha_{\text{water}}$   & 2 & 2 & 2     & $1/\text{Hz m}$   \\ \hline
$\alpha_{\text{concrete}}$& 30 & 30 & 30         & $1/\text{Hz m}$ \\ \hline
$\alpha_{\text{granite}}$ & - & - & 88         & $1/\text{Hz m}$ \\ \hline
\hline
\multicolumn{5}{|c|}{\textbf{Prior Model Parameters}} \\
\hline
\textbf{Parameter}   & \textbf{CC-KWave} & \textbf{CC-Exp} & \textbf{GB-Exp}     & \textbf{Unit} \\ \hline
Num. iterations & 100 & 100 & 100               & -             \\ \hline
$\sigma$             & 0.1 & 0.12 & 0.2 & Pascal        \\ \hline
p                    & 1.1 & 1.1 & $1.1$             & -             \\ \hline
q                    & 2.0 & 2.0 & $2.0$             & -             \\ \hline
T                    & 0.01 & 1 & 0.001 & -             \\ \hline
$\sigma_0$           & 2 & 2 & 2    &  $m^{-2}$\\ \hline
$\nu$                  & 10 & 10 & 5              & -             \\ \hline
$a$                  & 2 & 2 & 3                 & -             \\ \hline
\end{tabular}
\end{center}
\end{table}

Our forward model can account for the the shape of the collimated beam as discussed in \eqref{eq:apodFunc}. 
Therefore, we have to pick the value of the $\beta$ in \eqref{eq:apodFunc} that best matches the type of profile used in the experimental data. 
In Figure~\ref{fig:beams}, we visualize the effects of $\beta$ on beam spread and compare it with measured data.
 In Figure~\ref{fig:beams}(a-c), we plot  $\phi(v)^{(\beta)}e^{-\alpha_0 r(v)}$ for $\beta$ = 1, 4, and 8,  where $\phi(v)^{(\beta)}$ is the apodization function from \eqref{eq:apodFunc}, $\alpha_0$ is the attenuation coefficient in $m^{-1}$, $\theta(v)$ is the angle between beam direction and $v$, and $r(v)$ is distance from source to $v$.  
Increasing $\beta$ decreases the beam spread and makes it more collimated, with good perceptual fit to the measured data in Fig. \ref{fig:beams}(d) when $\beta = 8$.

\begin{figure*}[htb!]
\centering
\begin{tabular}{ccccc}
\tabularnewline 
\mbox{\hspace{0.25cm}} 
\includegraphics[width=2.5cm, height=4cm]{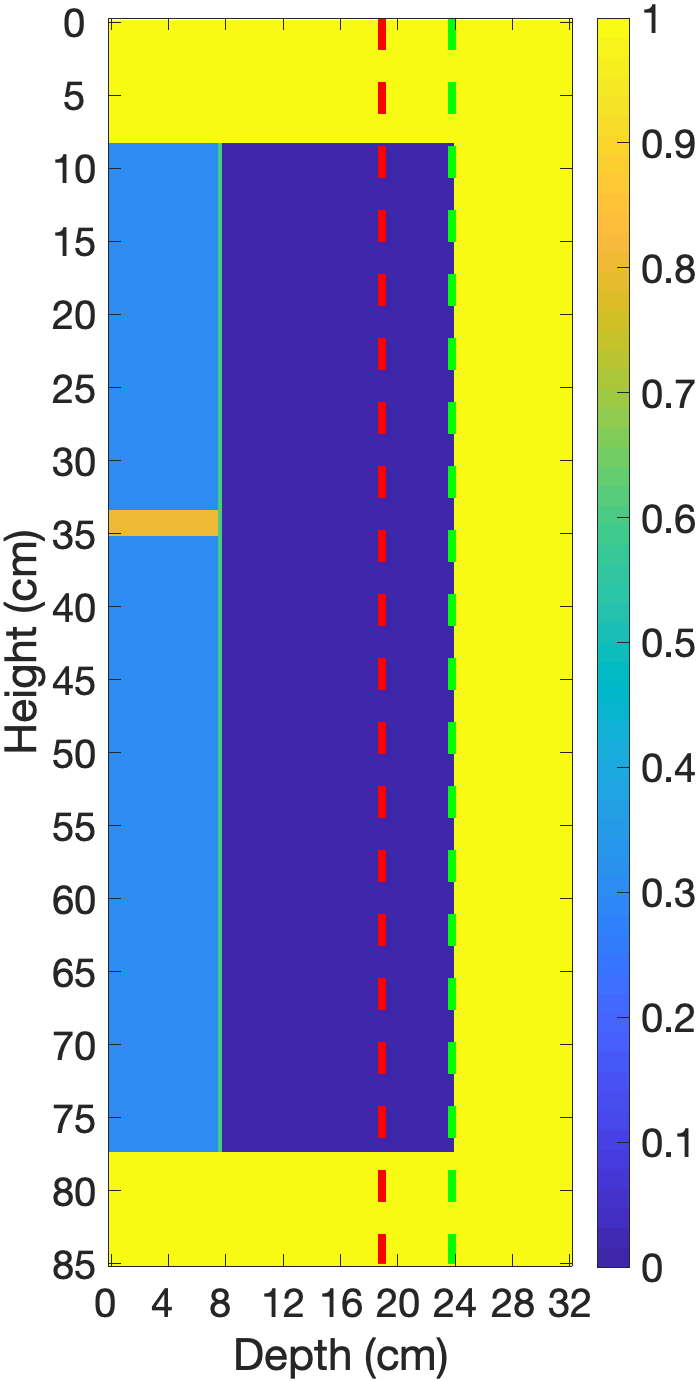} &
\mbox{\hspace{0.3cm}} 
\includegraphics[width=2.5cm, height=3.8cm]{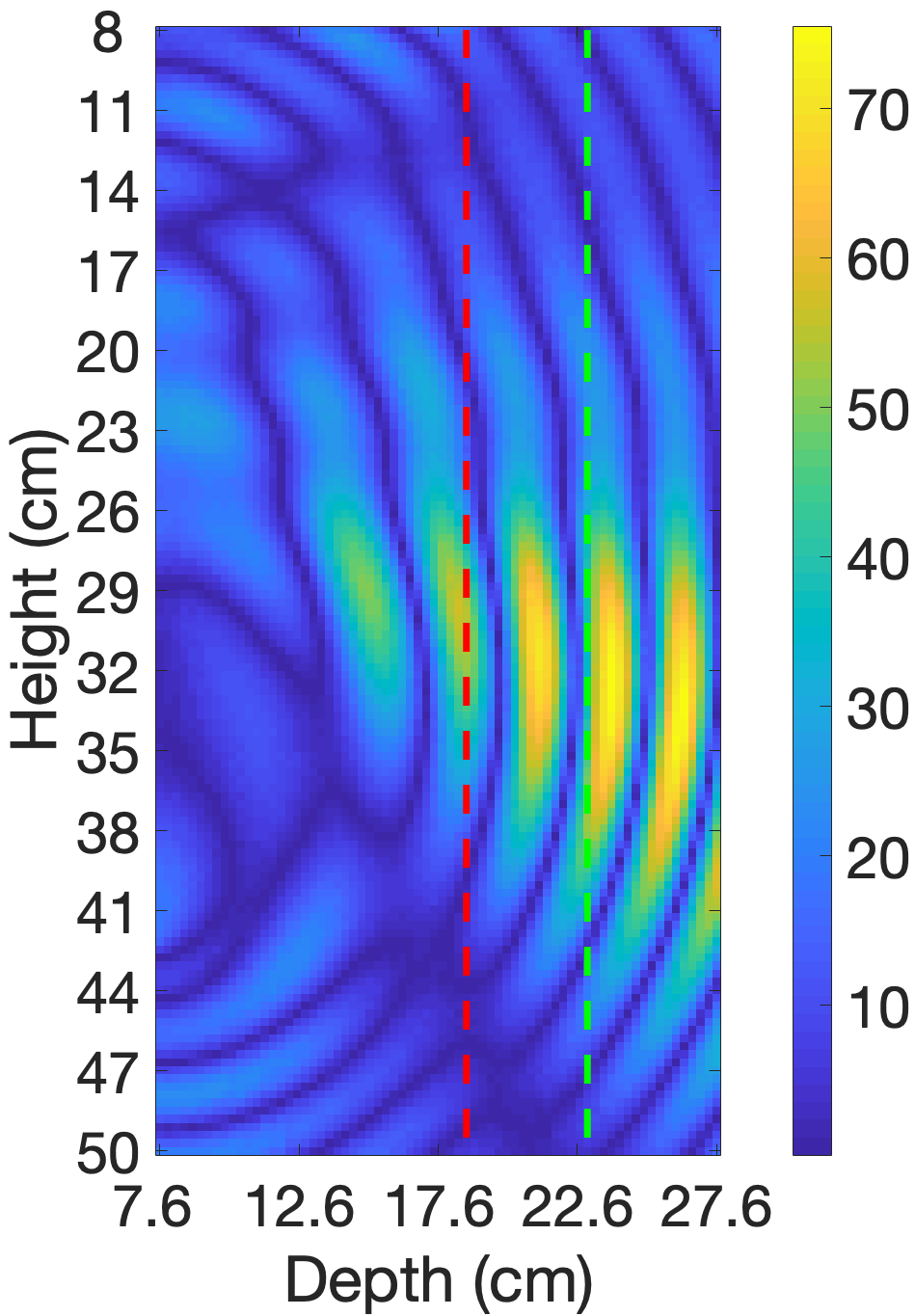} &
\mbox{\hspace{0.3cm}}
\includegraphics[width=2.5cm, height=4cm]{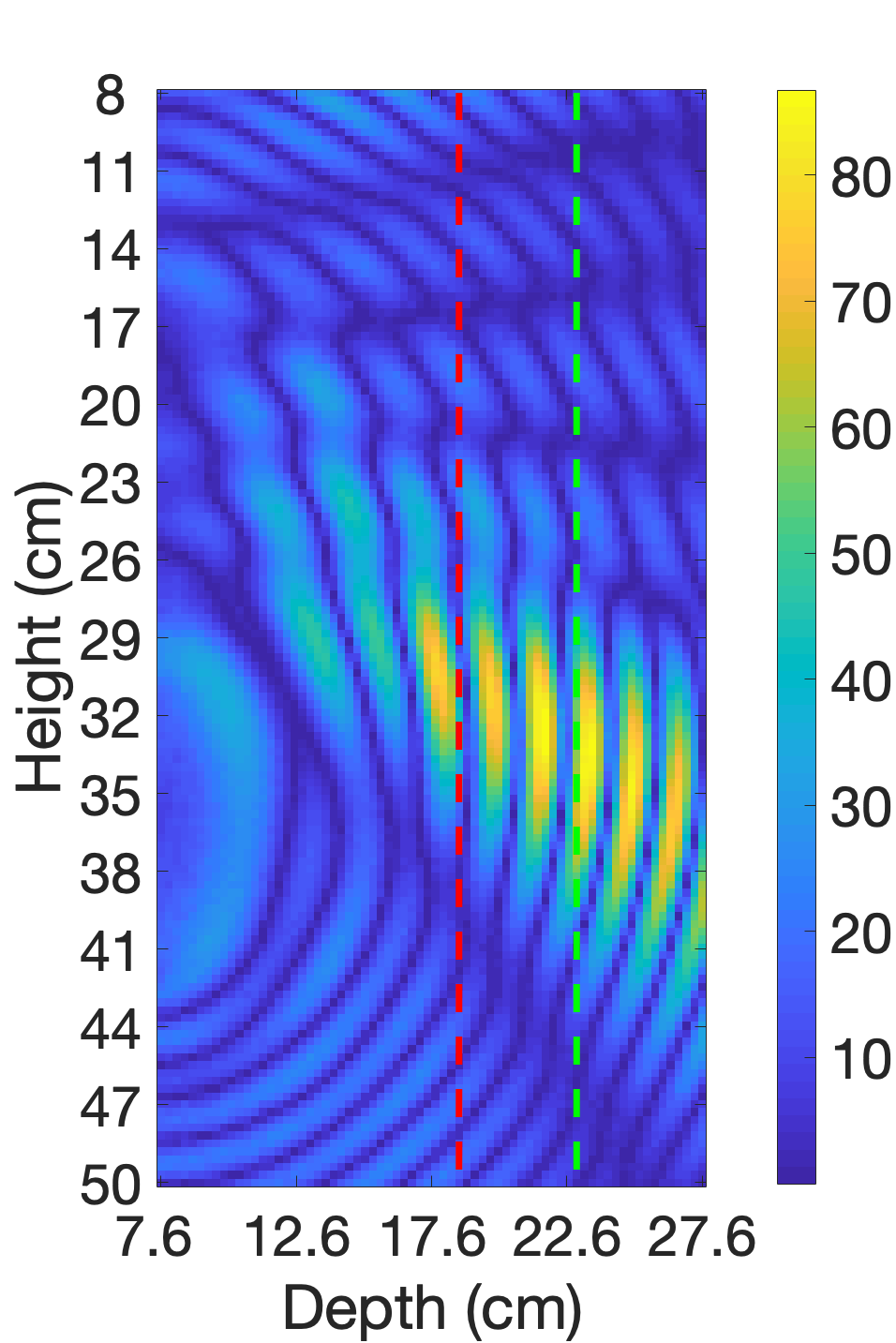}&
\mbox{\hspace{0.3cm}}
\includegraphics[width=2.5cm, height=4cm]{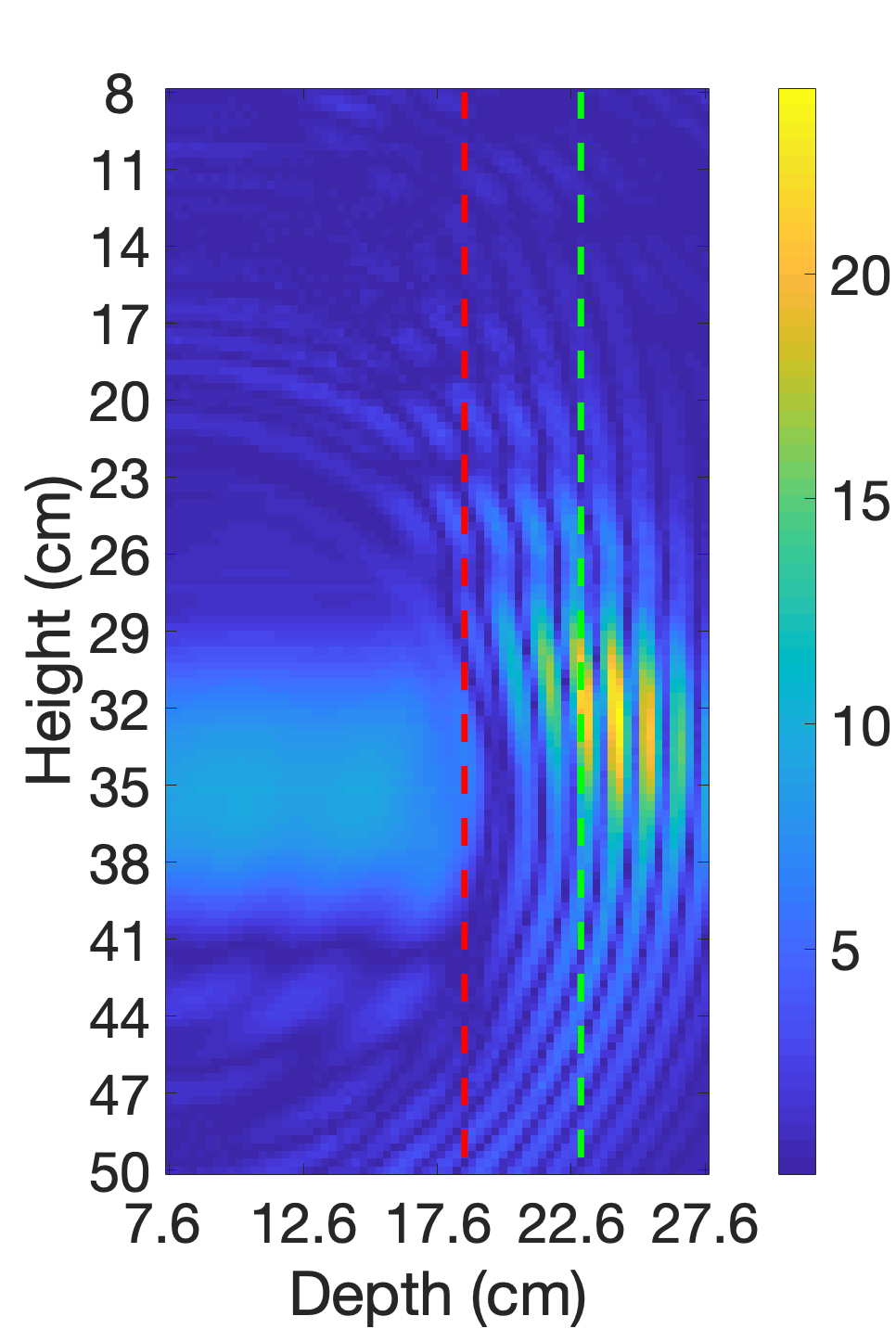}&
\mbox{\hspace{0.3cm}}
\includegraphics[width=2.5cm, height=4cm]{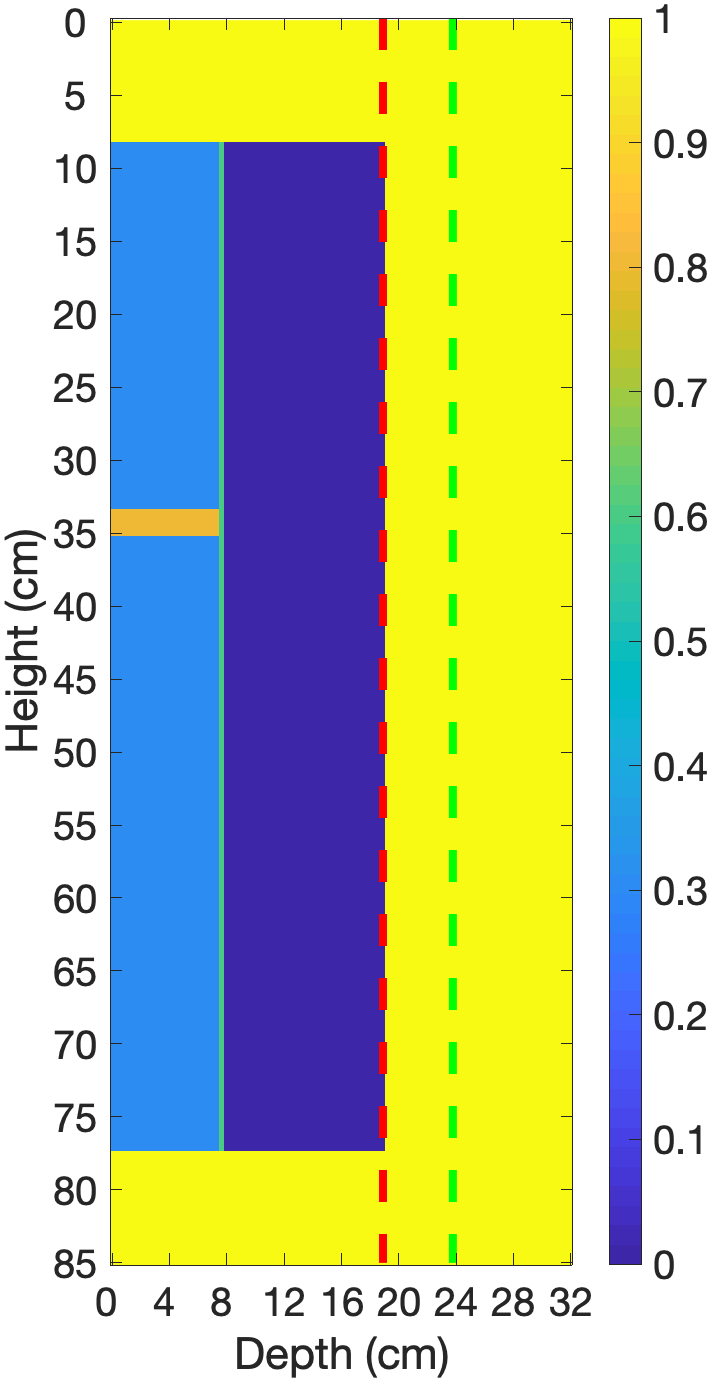}
 \tabularnewline
  \quad (a) GT & (b) SAFT 29kHz & (c) SAFT 42kHz & (d) SAFT 58kHz & (e) GT
  
\tabularnewline
\mbox{\hspace{8pt}}
\includegraphics[width=2.5cm, height=4cm]{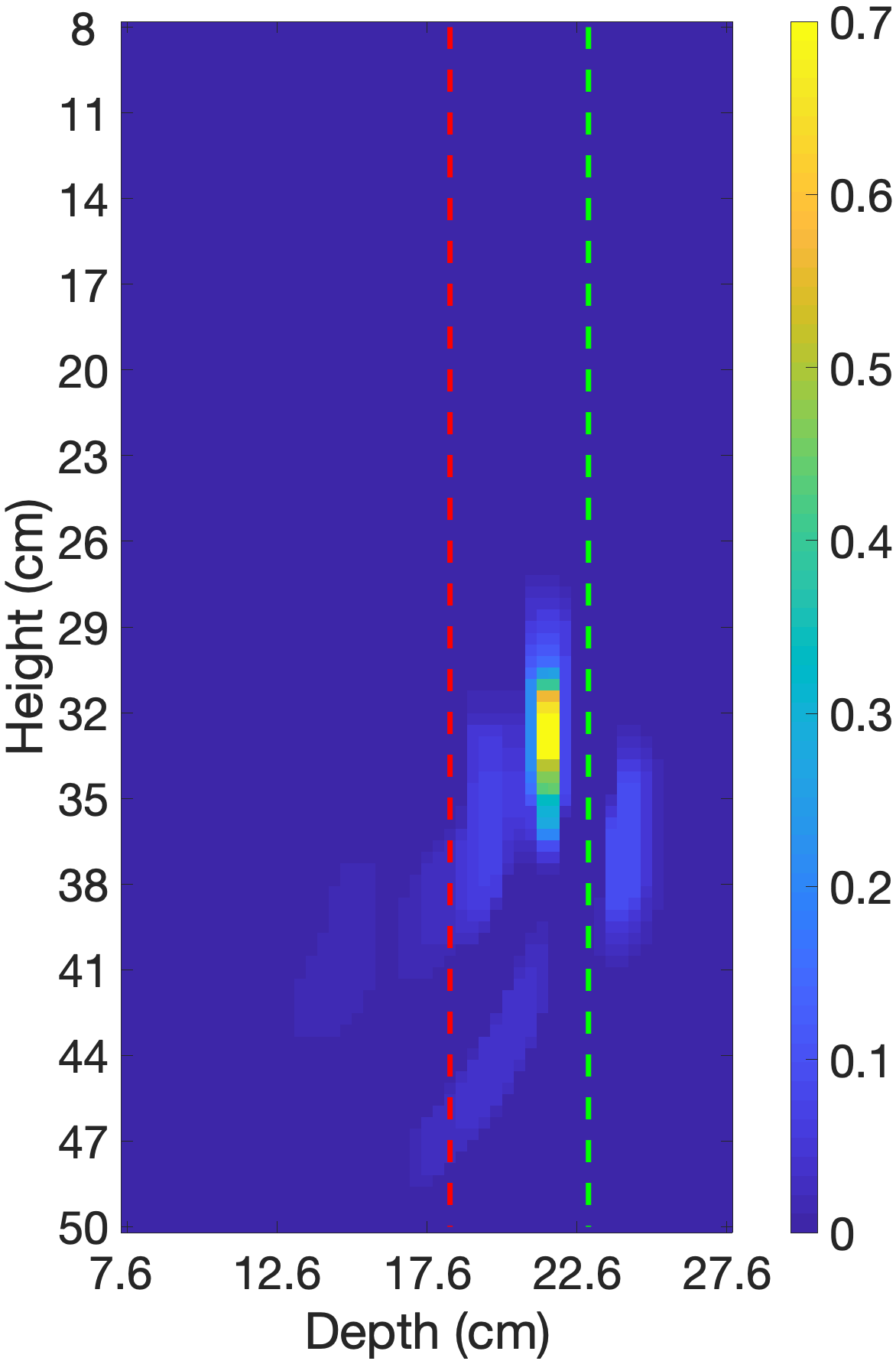}&
\mbox{\hspace{8pt}}
\includegraphics[width=2.5cm, height=4cm]{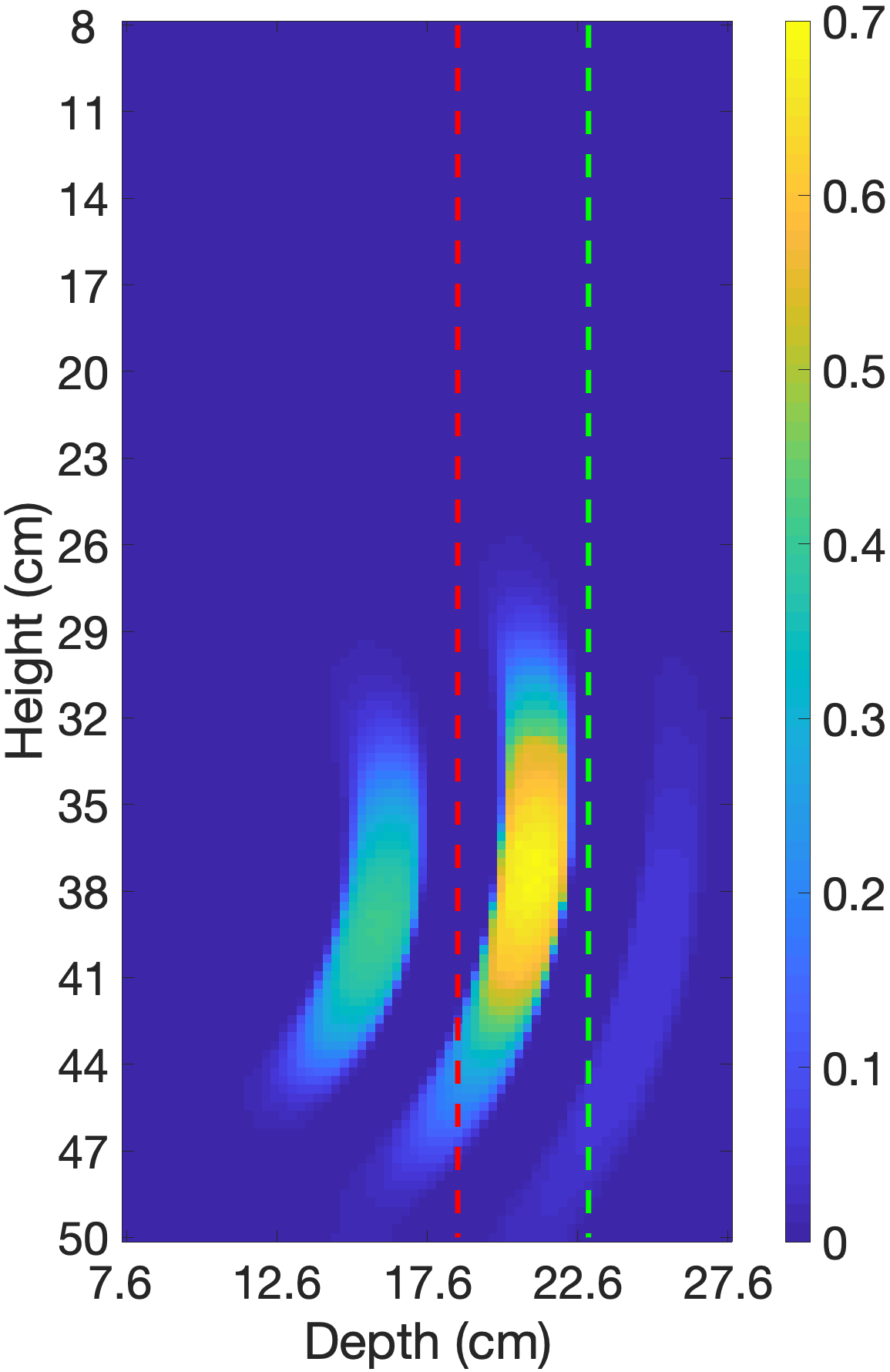}&
\mbox{\hspace{8pt}}
\includegraphics[width=2.5cm, height=4cm]{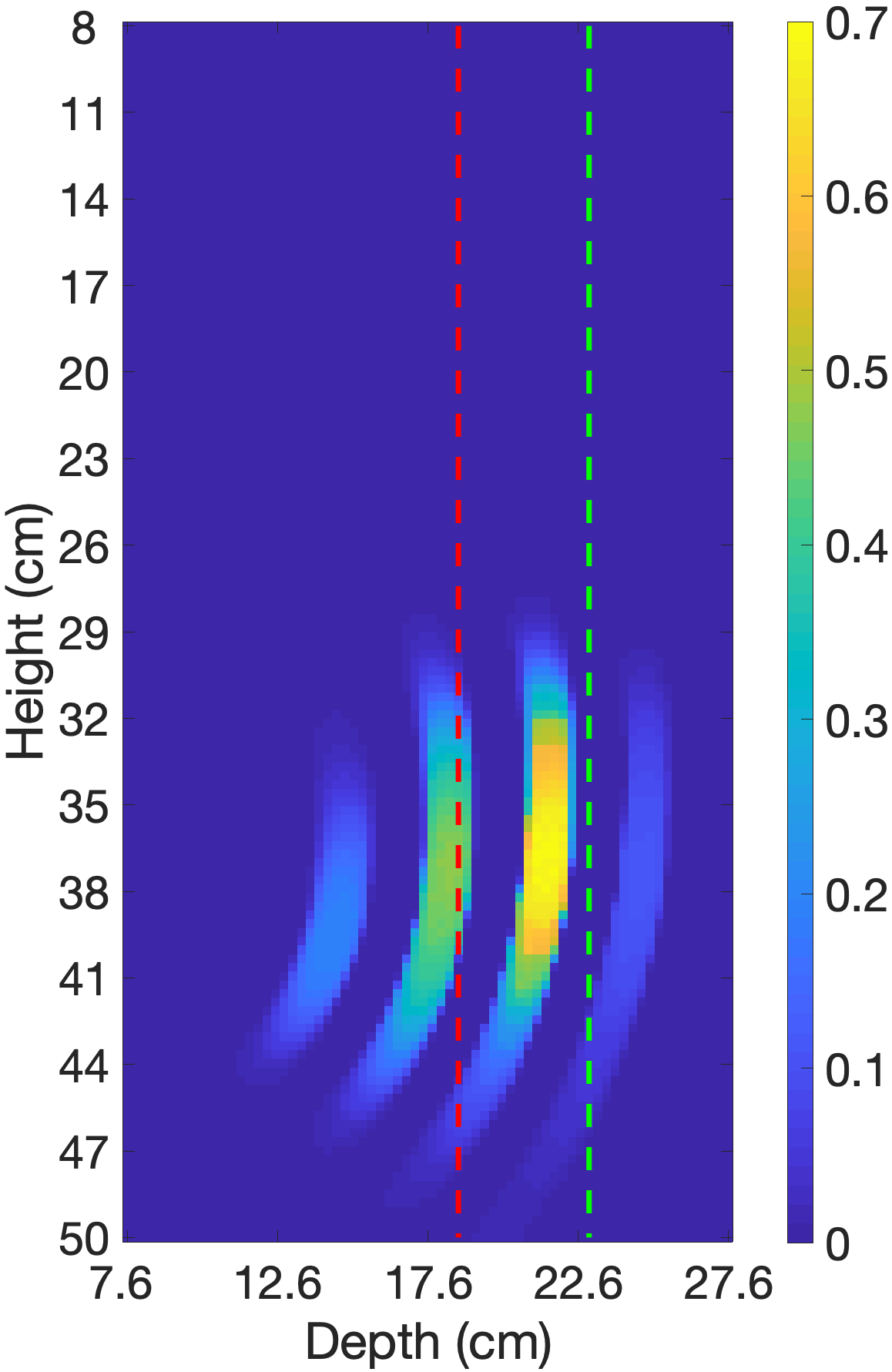}&
\mbox{\hspace{8pt}}
\includegraphics[width=2.5cm, height=4cm]{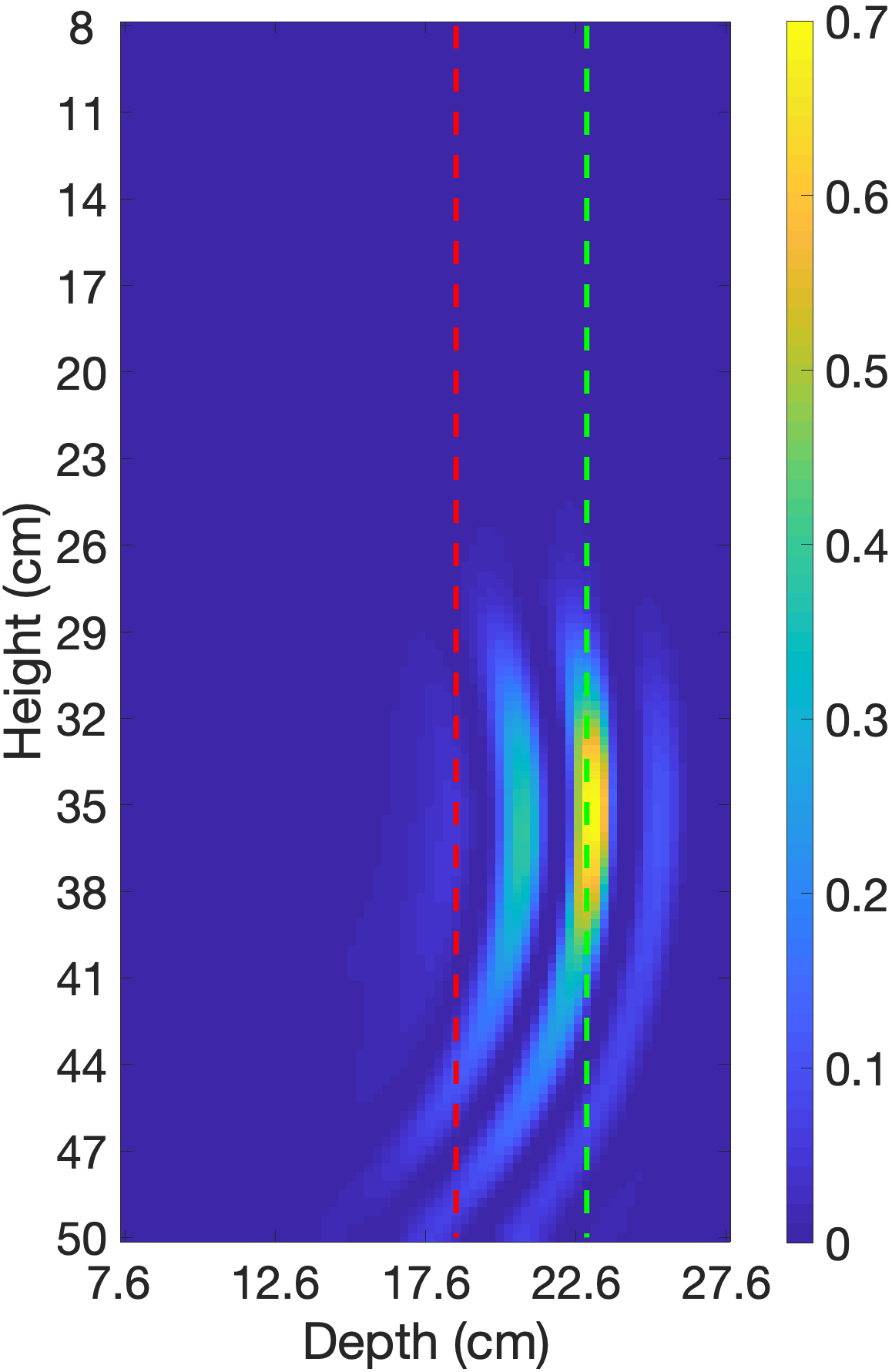} &
\mbox{\hspace{8pt}}
\includegraphics[width=2.5cm, height=4cm]{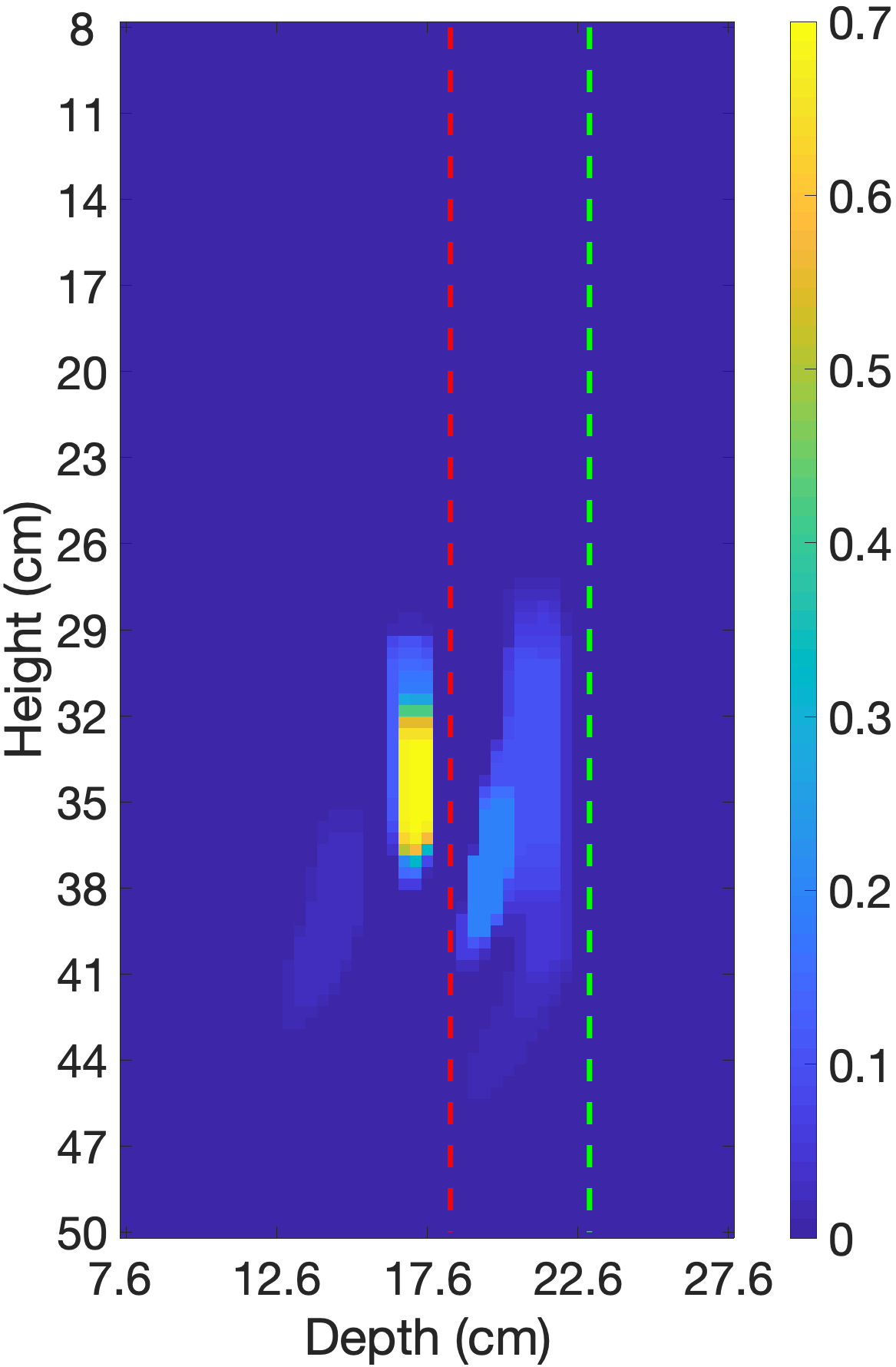}
 \tabularnewline
  \quad (f) MF-UMBIR & (g) UMBIR 29kHz & (h) UMBIR 42kHz & (i) UMBIR 58kHz & (j) MF-UMBIR
 \end{tabular}
\caption{
Results using synthetic data generated with K-Wave for concrete cylinder experiment.
(a) ground truth without notch;
(b), (c), and (d) single frequency SAFT reconstructions at 29 kHz, 42 kHz, and 58 kHz without notch;
(e) ground truth with notch;
(f) multi-frequency UMBIR reconstruction of K-Wave data without notch;
(g), (h), and (i) single frequency UMBIR reconstructions at 29 kHz, 42 kHz, and 58 kHz without notch;
(j) multi-frequency UBMIR reconstruction of K-Wave data with notch.
The red and green dashed lines indicate the notch and backwall locations, respectively.
Notice that the two MF-UMBIR reconstructions of (f) and (j) accurately reconstruct the location of the notch and the back wall.
} 
\label{fig:Concrete-Cylinder-KWave}
\end{figure*}

\subsection{Concrete Cylinder Experiment}
\label{LANL-I_Experiment}

Figure~\ref{fig:Concrete-Cylinder-Experiment} illustrates the experimental setup for the concrete cylinder experiment that was performed at LANL \cite{pantea2019collimated}. 
The concrete cylinder is designed to represent a concrete well bore with a central hole that contains the acoustic imaging sensor.

Figure~\ref{fig:Concrete-Cylinder-Experiment}(a) shows that on one side of the concrete cylinder there is a notch marked with red paint that is 50 mm deep and subtends an angle of approximately $45^{\circ}$.
Figure~\ref{fig:Concrete-Cylinder-Experiment}(c) is a detailed diagram showing the dimensions and positions of all the components. Notice that the acoustic imaging sensor is placed in the center of the cylinder with the receiver array at the top and the acoustic transmitter at the bottom.
The blue box in Figure~\ref{fig:Concrete-Cylinder-Experiment}(c) shows the region in which the K-Wave simulation is performed for the generation of synthetic data, and the red box shows the region in which the UMBIR reconstruction is computed. 

Figure~\ref{fig:Concrete-Cylinder-Experiment}(b) shows the receiver array along with the collimated transmitter hanging below.
This entire assembly is positioned inside the bore hole at the center of the concrete cylinder where it can be rotated by the computer-controlled rotation system.
Figure~\ref{fig:Concrete-Cylinder-Experiment}(d) is a diagram showing the position of the transducers in the sensor array. 
We note that the transmitters and receivers are immersed in water to facilitate acoustic transmission into and out of the concrete. 
However, there is an vacuum barrier shown as a green line in Figure~\ref{fig:Concrete-Cylinder-Experiment}(c) that is positioned between the transmitter and receivers aiming to block the direct arrival signal.

Table~\ref{table:TransmitterParams} lists out the transmission parameters at the three different excitation frequencies that were used.
The data was collected at $5^\circ$ increments using a rotational span of $180^\circ$, with the sensor assembly facing the middle of the notch at the rotational position of $90^\circ$.

\begin{figure*}[htb]
\centering
\begin{tabular}{cccc}
\small{29 kHz} & \small{42.4 kHz} & \small{58 kHz} & \small{MF-UMBIR}\\
\put(0,0)\centering{\rotatebox{90}{{\quad \quad  $0^\circ$ (no notch)}}}   
\includegraphics[width=0.15\textwidth]{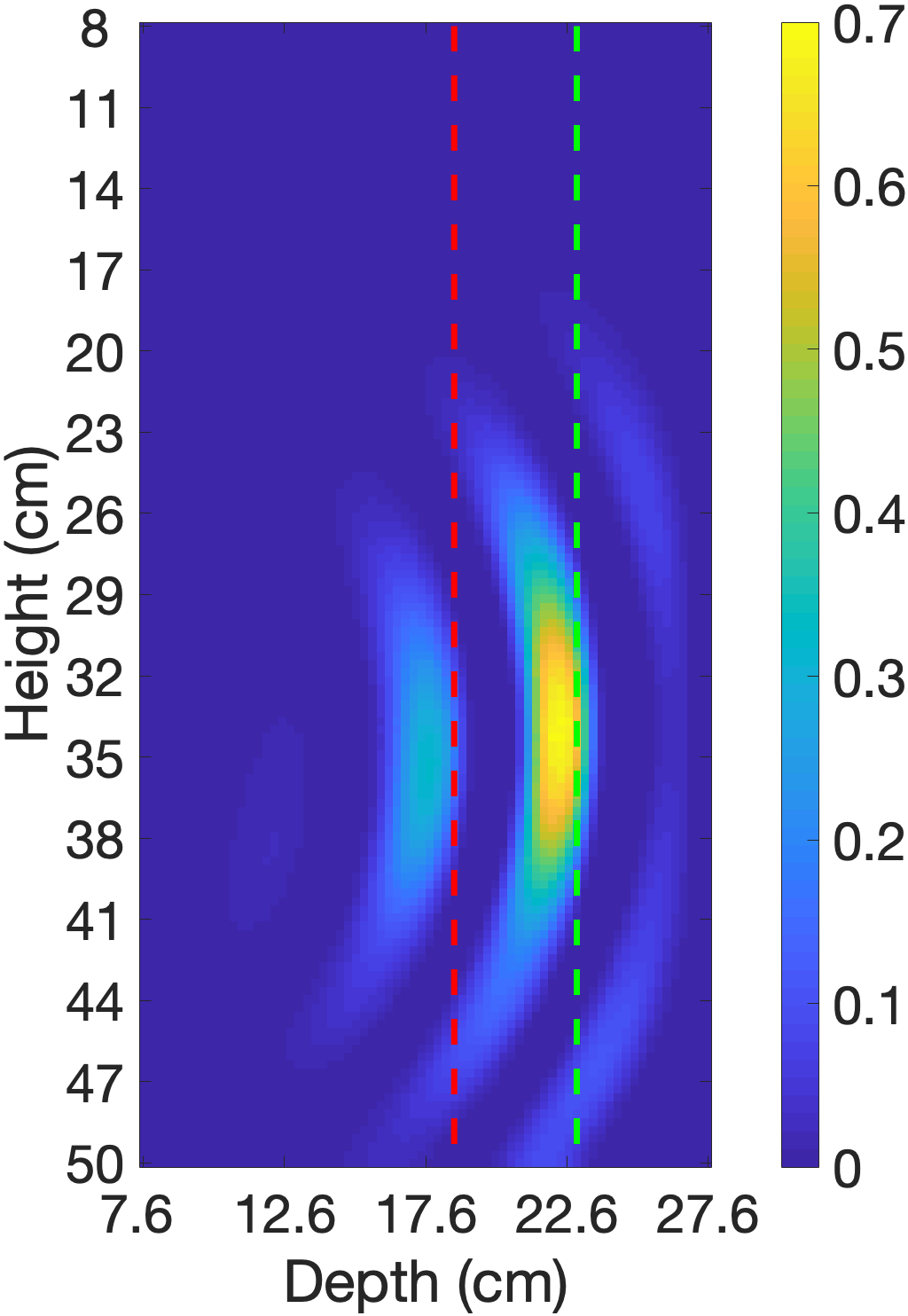} &
\includegraphics[width=0.15\textwidth]{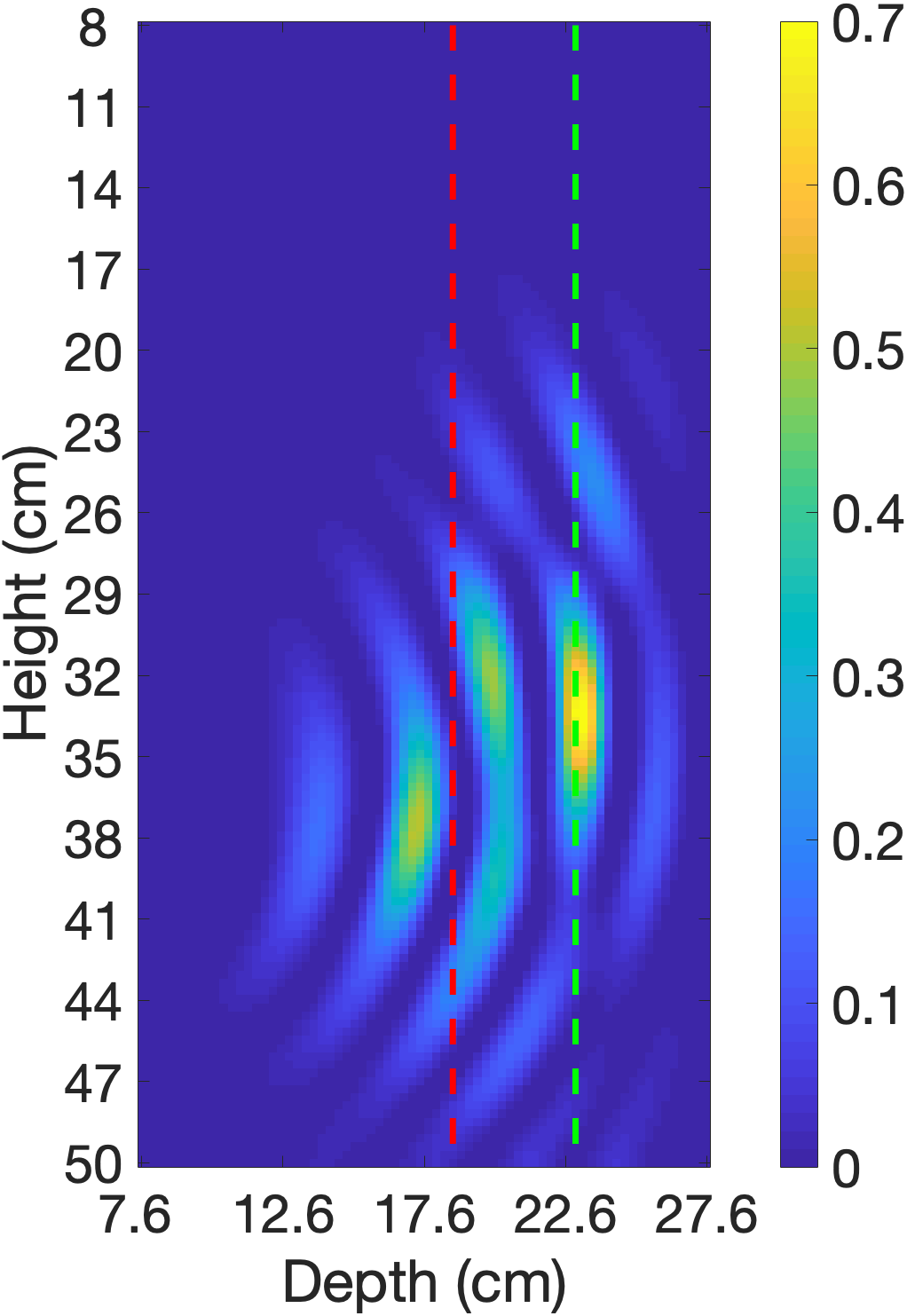} &
\includegraphics[width=0.15\textwidth]{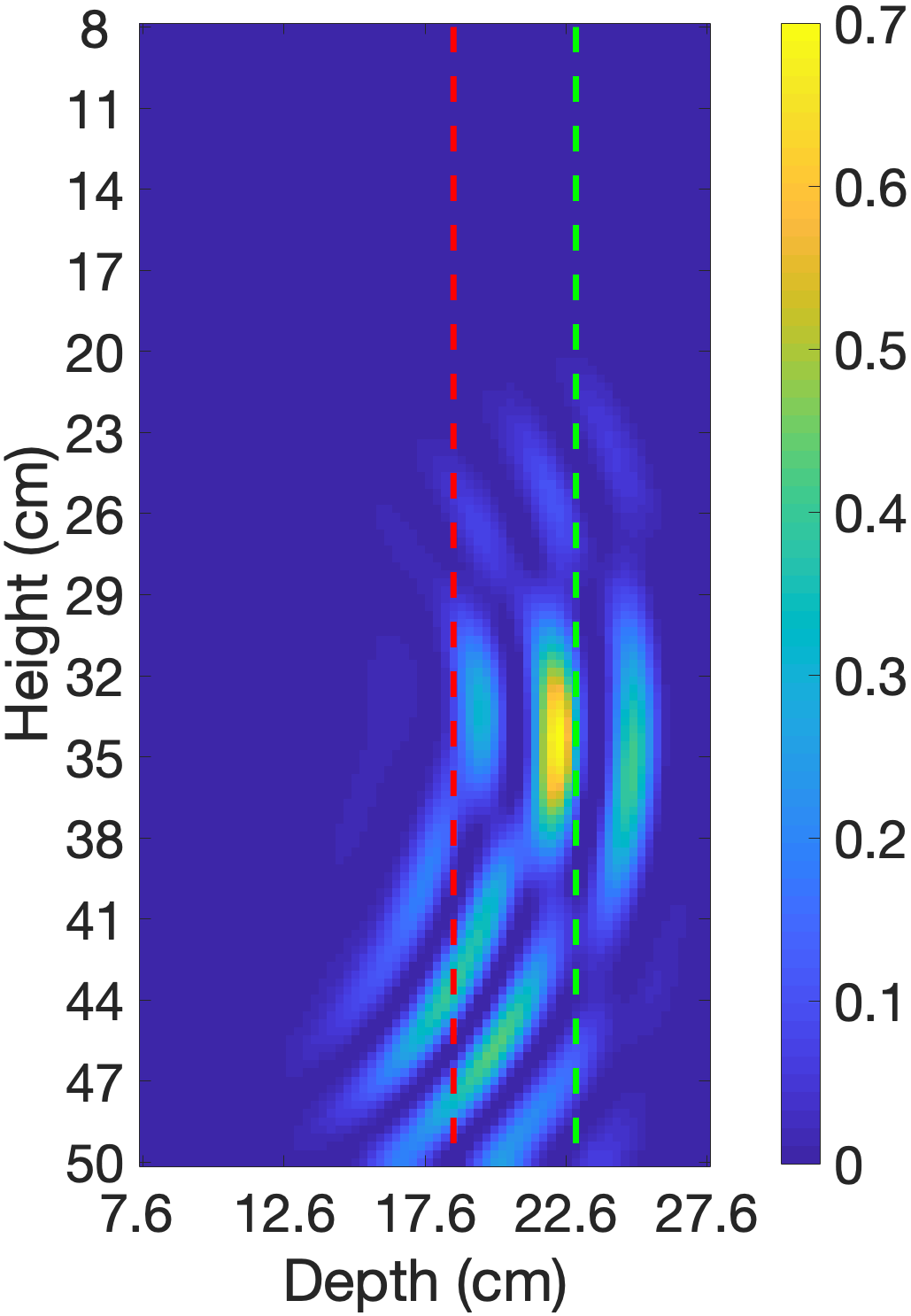} &
\includegraphics[width=0.15\textwidth]{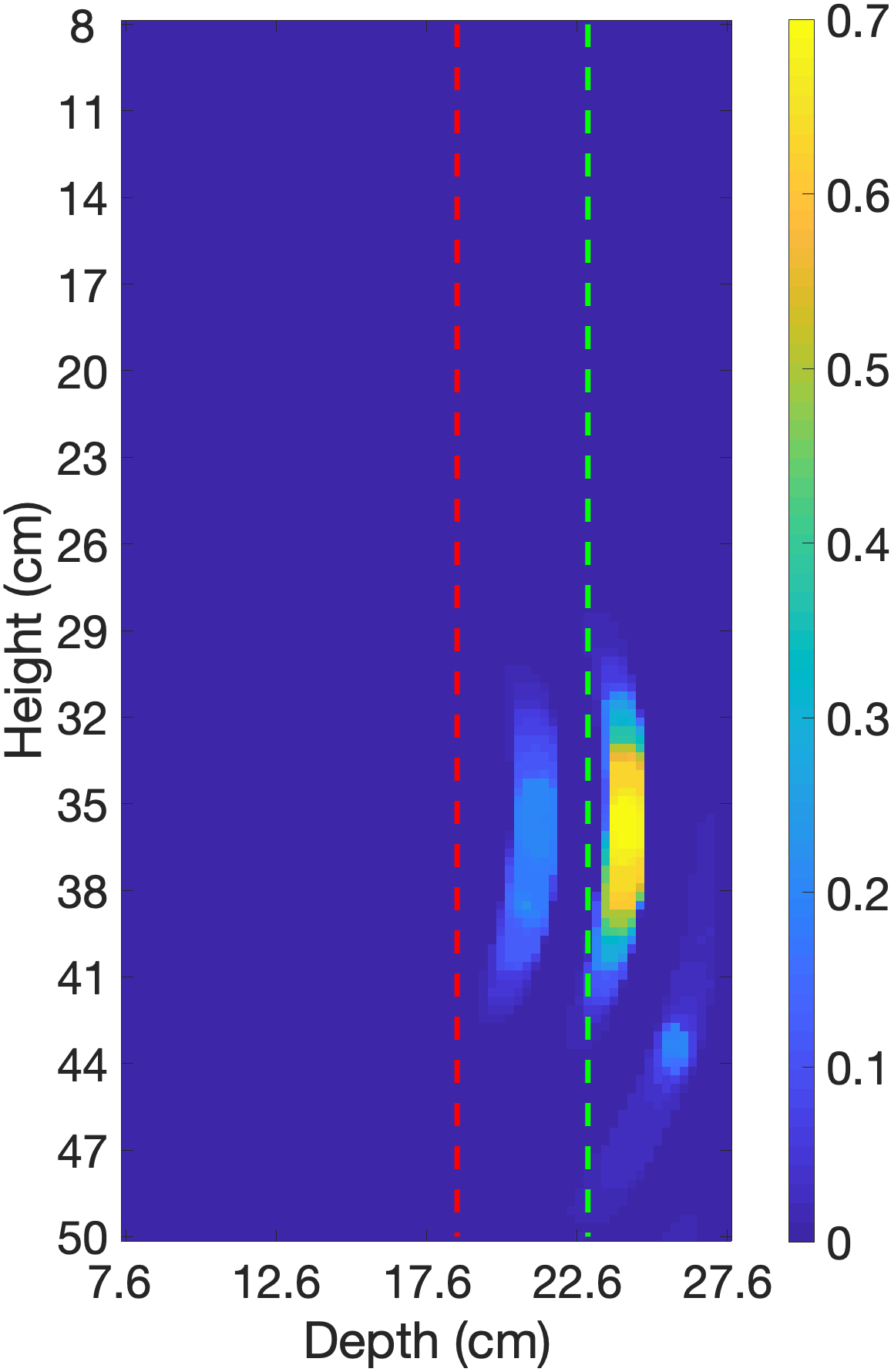} \\
\put(0,0)\centering{\rotatebox{90}{{\quad \quad  $90^\circ$ (notch)}}}   
\includegraphics[width=0.15\textwidth]{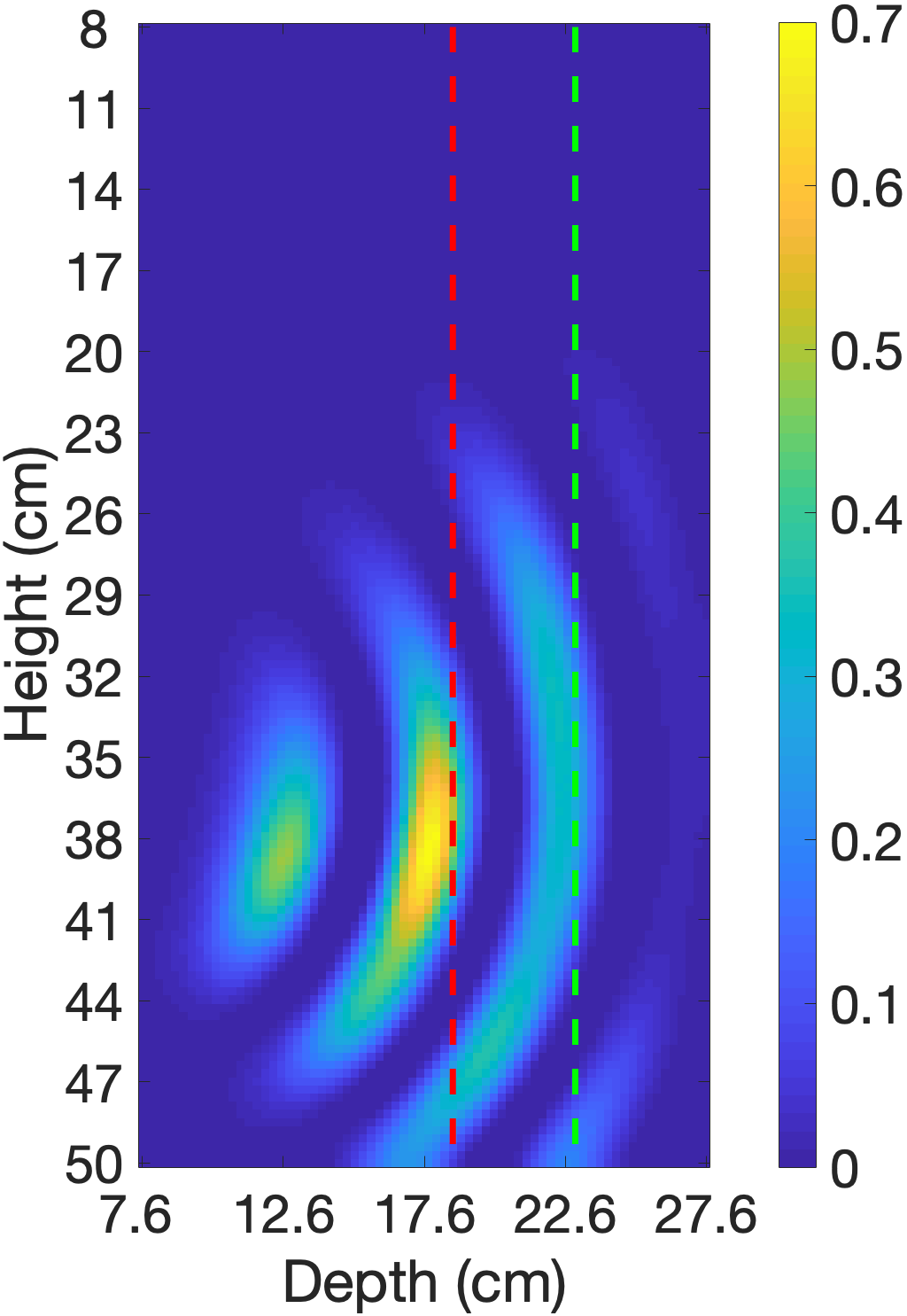} &
\includegraphics[width=0.15\textwidth]{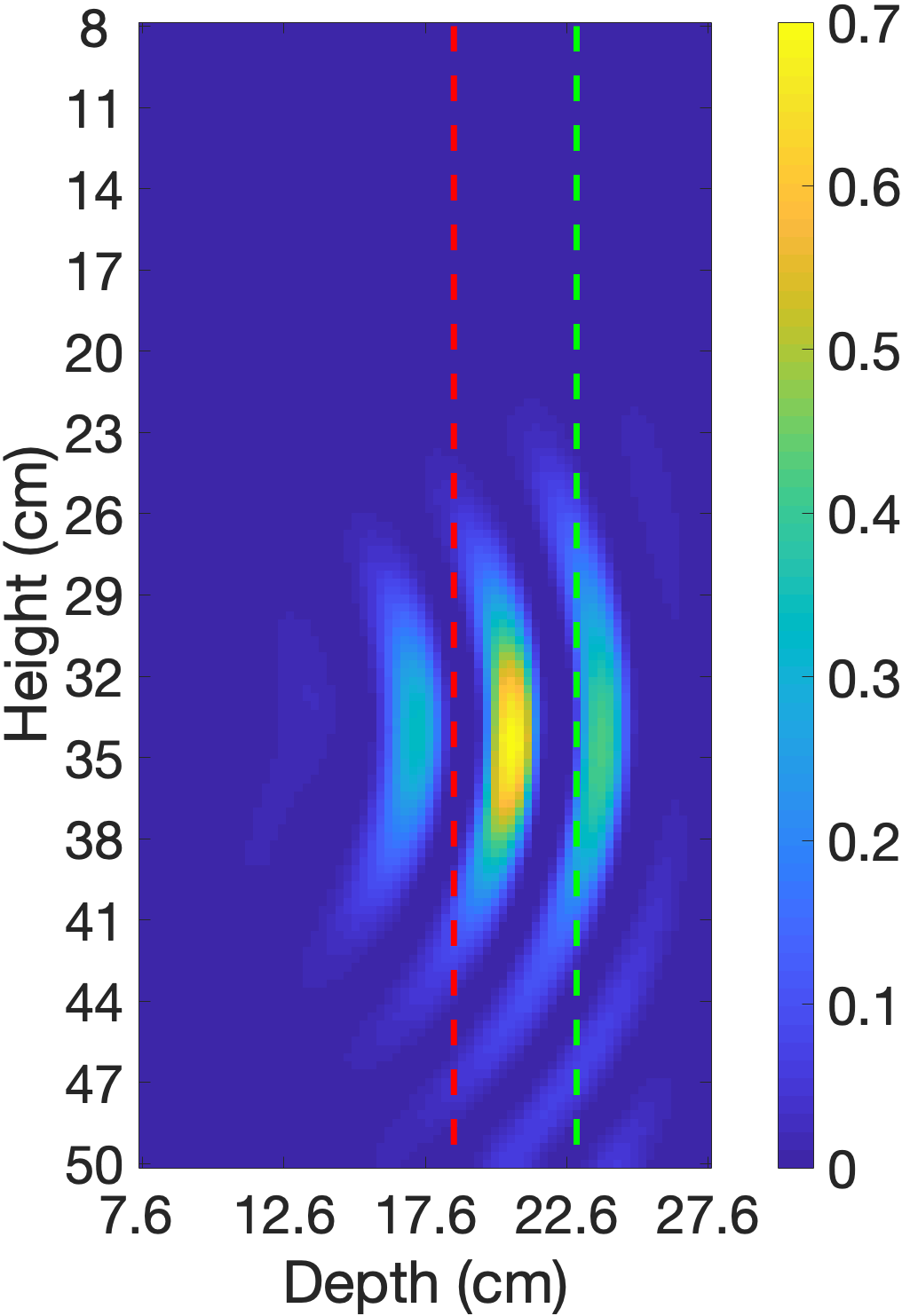} &
\includegraphics[width=0.15\textwidth]{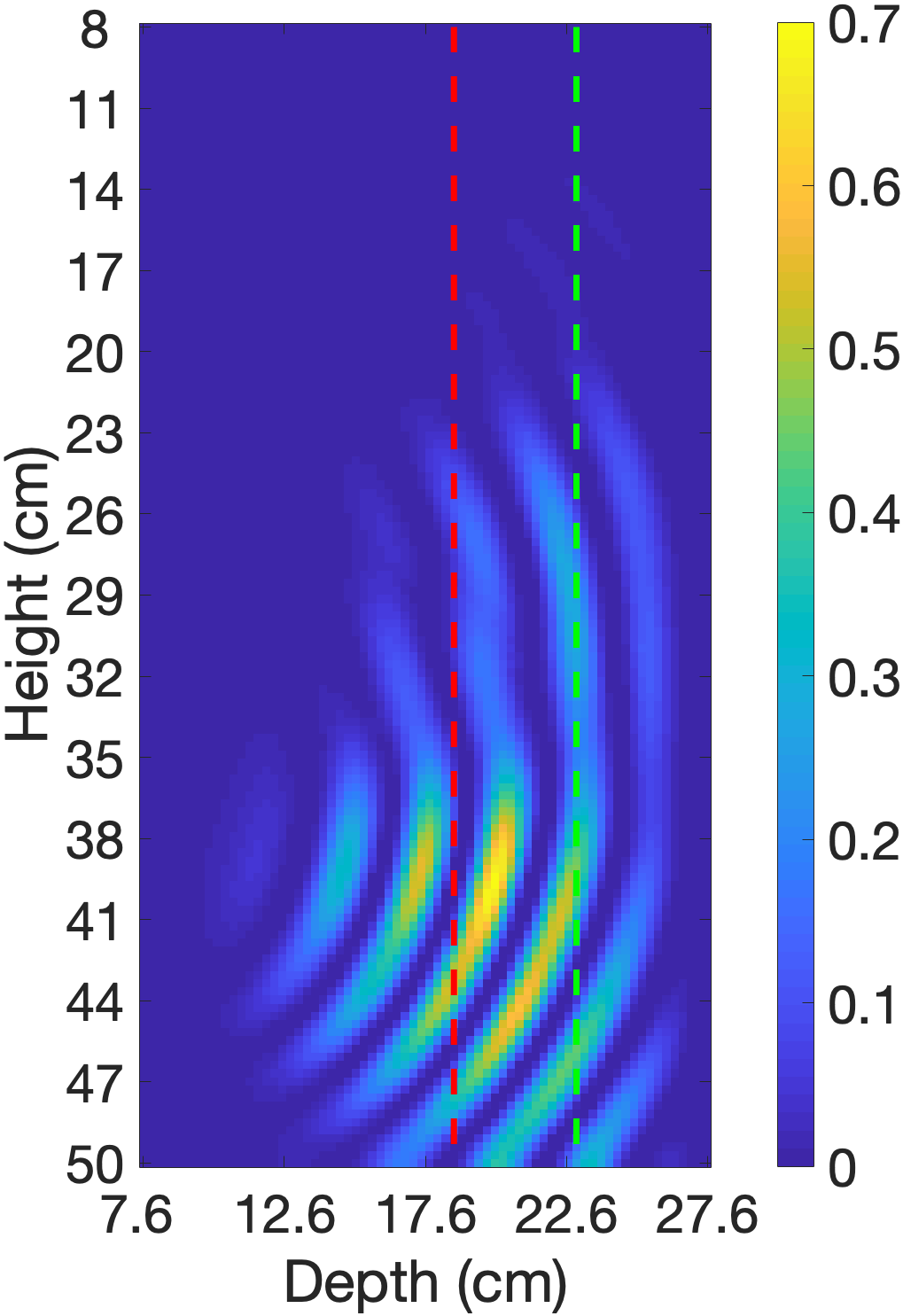} &
\includegraphics[width=0.15\textwidth]{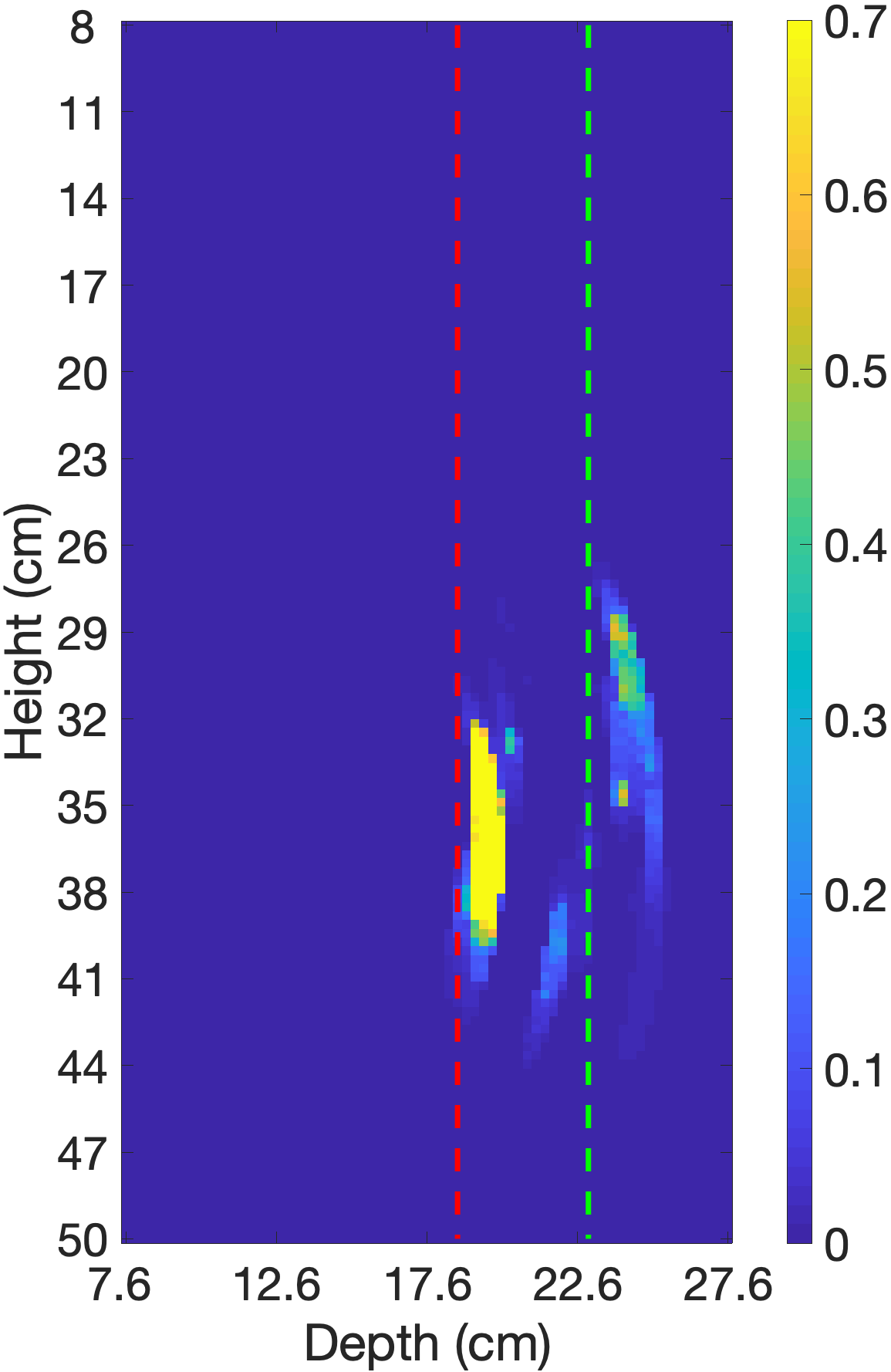}
\end{tabular}
\caption{Results of reconstructing experimental data collected from concrete cylinder.
Column 1 to 3 of figure: Single frequency UMBIR reconstructions at excitation frequencies of 29 kHz, 42.4 kHz, 58 kHz and multi-frequency UMBIR, without and with the notch.
Column 4 of figure: Multi-frequency UMBIR reconstruction without and with the notch.
Notice that the multi-frequency UMBIR reconstruction results in much better localization and greater accuracy of the back wall location.
}
\label{fig:Concrete-Cylinder-Real}
\end{figure*}

\begin{table}[htb]
\caption{Transmit signal parameters used for the concrete cylinder experiment.}
\label{table:TransmitterParams}
\centering
\setlength\tabcolsep{5pt}
\begin{tabular}{|c|c|c|} 
\hline
Excitation Frequency & Duration & Pulse Shape \\ \hline \hline
29 kHz  & 200 $\mu$s & Tukey \\  \hline
42.4 kHz  & 200 $\mu$s & Tukey \\  \hline
58 kHz  & 50 $\mu$s & Tukey \\  \hline
\end{tabular}
\end{table}

\subsubsection{K-Wave Results}

\begin{table}[htb]
\caption{K-Wave parameters used for concrete cylinder.}
\label{table:GeoParameters}
\centering
\setlength\tabcolsep{5pt}
\begin{tabular}{|c|c|c|} 
\hline
Component & Dimensions & Units \\ \hline \hline
Computational domain height  & 855 & mm \\  \hline
Computational domain width  & 362 & mm \\ \hline
Pixel pitch in x/y direction & 3 & mm \\ \hline
perfectly matched layer (PML) size & 20 & samples \\ \hline
water acoustic speed & 1.5 & km/s \\ \hline
Plexiglas acoustic speed & 2.82 & km/s \\ \hline
concrete acoustic speed & 2.62 & km/s \\ \hline
water density & 997 & kg/m$^3$ \\ \hline
Plexiglas density & 1180 & kg/m$^3$ \\ \hline
concrete density & 1970 & kg/m$^3$ \\ \hline
\end{tabular}
\end{table}

In order to better understand the concrete cylinder experiment, we first perform reconstructions using synthetic data generated with the K-Wave simulation using the parameters shown in Table~\ref{table:GeoParameters}. 
Table~\ref{table:UMBIR-Reconstruction-Params} lists the parameter settings used in K-Wave data reconstructions under CC-KWave, and all simulations are done using a perfectly matched layer (PML) that extends in all directions from the outer boundary of the computational domain. 
The regions beyond the top, bottom, and back wall are assumed to be air (yellow region) to mimic the real data, and an isolating layer (void) was placed between the source and receivers to partially block direct arrival signals. 

Figure~\ref{fig:Concrete-Cylinder-KWave} shows a comparison of single frequency UMBIR, multi-frequency UMBIR, and SAFT when reconstructing synthetically generated concrete cylinder data.
Figure~\ref{fig:Concrete-Cylinder-KWave}(a) and (e) show the ground truth used to generate the data with and without the notch, respectively.
Without the notch, the back wall should be located at a depth of 18.85 cm (shown with a red dotted line), and with the notch it should be located at a depth of 23.85 cm (shown with a green dotted line).

The single frequency SAFT reconstructions in Figure~\ref{fig:Concrete-Cylinder-KWave}(b-d) show multiple reflections from the back wall and other artifacts at each excitation frequency, which leads to uncertainty in the estimated location of the back wall. 
In contrast, the single frequency UMBIR reconstructions in Figure~\ref{fig:Concrete-Cylinder-KWave}(g-i) show that even single-frequency UMBIR reconstructions provide fewer artifacts and more accurate localization of the back wall than SAFT.

The multi-frequency UMBIR reconstruction without the notch in Figure~\ref{fig:Concrete-Cylinder-KWave}(f) is a substantial improvement over the single frequency UMBIR reconstructions. 
The same is true for the multi-frequency UMBIR reconstruction with the notch in Figure~\ref{fig:Concrete-Cylinder-KWave}(j).
This demonstrates that joint processing of low and high excitation frequencies results in substantially better reconstruction quality than single-frequency reconstruction.

\subsubsection{Experimental Data Results}
\label{sec:LANL_I_RealData}

Figure~\ref{fig:Concrete-Cylinder-Real} depicts a selection of cross-section reconstruction results using the measured experimental data.
The left 3 columns of Figure~\ref{fig:Concrete-Cylinder-Real} show single frequency UMBIR cross-section reconstructions for the excitation frequencies 29 kHz, 42.4 kHz, 58 kHz without and with the notch.
Column 4 shows the multi-frequency UMBIR reconstruction without and with the notch.
We again note that the back wall without and with the notch are located at 23.85 cm and 18.85 cm, respectively.
So from this we see that the joint reconstruction of the multiple frequencies more accurately localizes of the back wall.

\begin{figure}[htb]
\begin{center}
\includegraphics[width=0.3\textwidth]{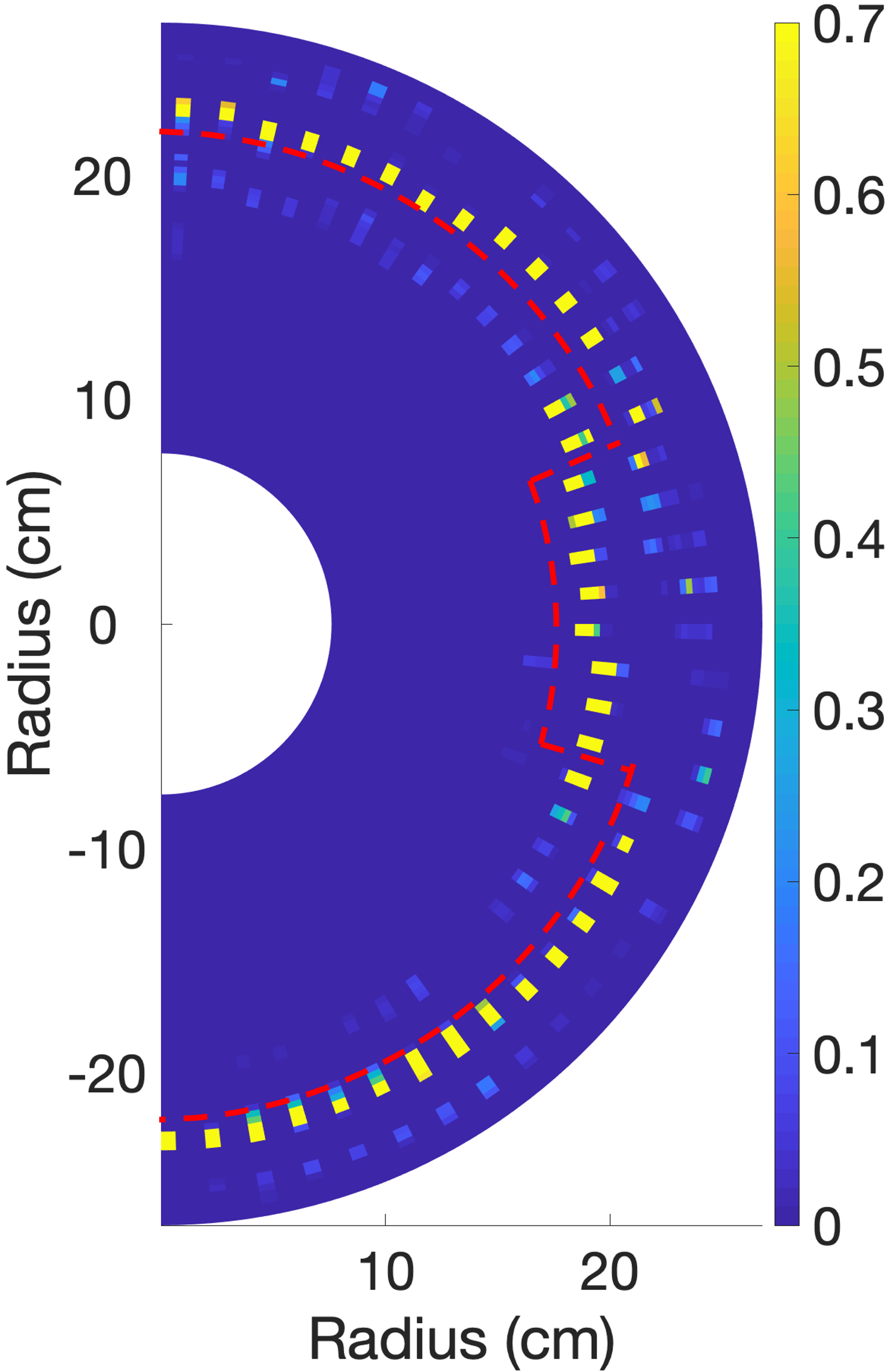}
\flushleft\caption{The multi-frequency UMBIR panoramic stitched reconstruction of the concrete cylinder at a fixed height of 27 cm obtained at each of 37 views in the range from $0^\circ$ to $180^\circ$ and displayed in polar coordinates. The dashed black square shows the outer edges of the GB, and the red circles indicate the two embedded defects. The red dashed line shows the location of the wall and notch, which is closely followed by MF-UMBIR reconstructions. }
\label{fig:Concrete-Cylinder-Real-Panoramic}
\end{center}
\end{figure}

The panoramic reconstruction view of the concrete cylinder in Figure~\ref{fig:Concrete-Cylinder-Real-Panoramic} provides a more visually intuitive interpretation of the multi-frequency UMBIR reconstruction.
The panoramic reconstruction is formed by combining the views from each measured angle (37 equi-spaced angles from $0^\circ$ to $180^\circ$) to form a 2-dimensional horizontal cross-section at a fixed height of 27 cm.
In this case, both the back wall with and without the notch is shown as a dotted red line. 
Notice that the multi-frequency reconstruction localization closely follows its true location.
There is some error in capturing the transition along the notch, but the locations are captured accurately over a large range of angles, and there are few reflection artifacts.

%
%

Table~\ref{table:A_TimeComp} gives approximate computation times for 100 iterations of single and multi-frequency UMBIR using an optimized code running on an Intel(R) Core(TM) i7 CPU E5-2603 0 @1.80 GHz, 32.00 GB RAM.
The table lists out both the time required to compute the system matrices, $A$ and $D$, and the time required to reconstruct one 2D image for the concrete cylinder data set shown in Figure~\ref{fig:Concrete-Cylinder-KWave}(i) and (j).

\begin{table}[htb]
\caption{Computation time to obtain the UMBIR and MF-UMBIR reconstructions in Figure~\ref{fig:Concrete-Cylinder-KWave}(i) and (j).}
\label{table:A_TimeComp}
\begin{tabular}{|l|l|l|} \hline
Computation Time & UMBIR & MF-UMBIR \\ \hline  
Time to compute system matrices & 26.6 s & 91.3 s \\ \hline 
Time to perform reconstruction & 44.3 s & 205.2 s \\ \hline 
\end{tabular}
\end{table}

\subsection{Granite Block Experiment}
\label{sec:GraniteBlockExperiment}

\begin{figure*}[htb]
\centering
\begin{tabular}{ccccc}
\tabularnewline
\includegraphics[width=0.21\textwidth]{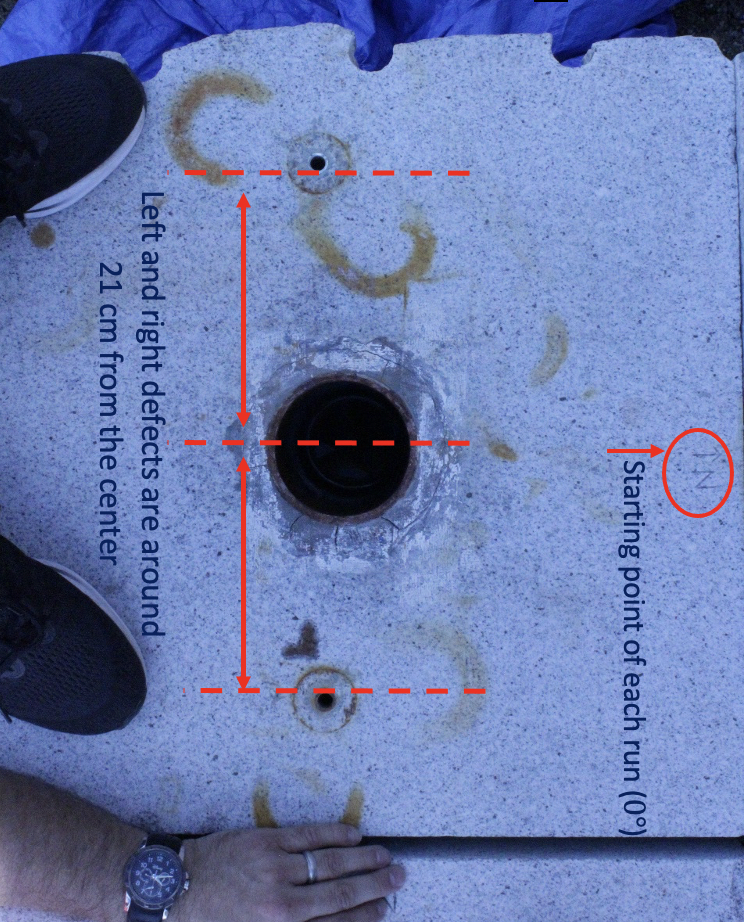} &
\includegraphics[width=0.15\textwidth]{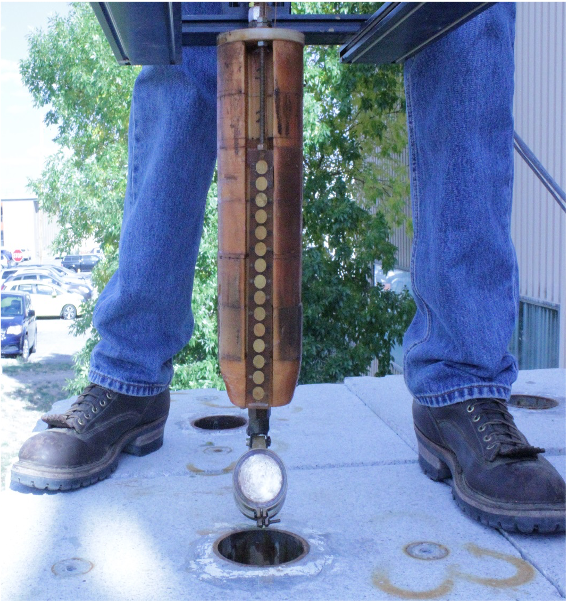} &
\includegraphics[width=0.18\textwidth]{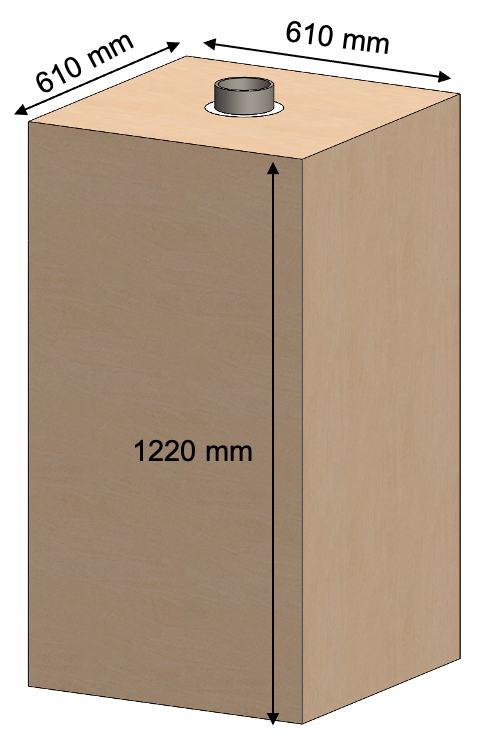} &
\includegraphics[width=0.25\textwidth]{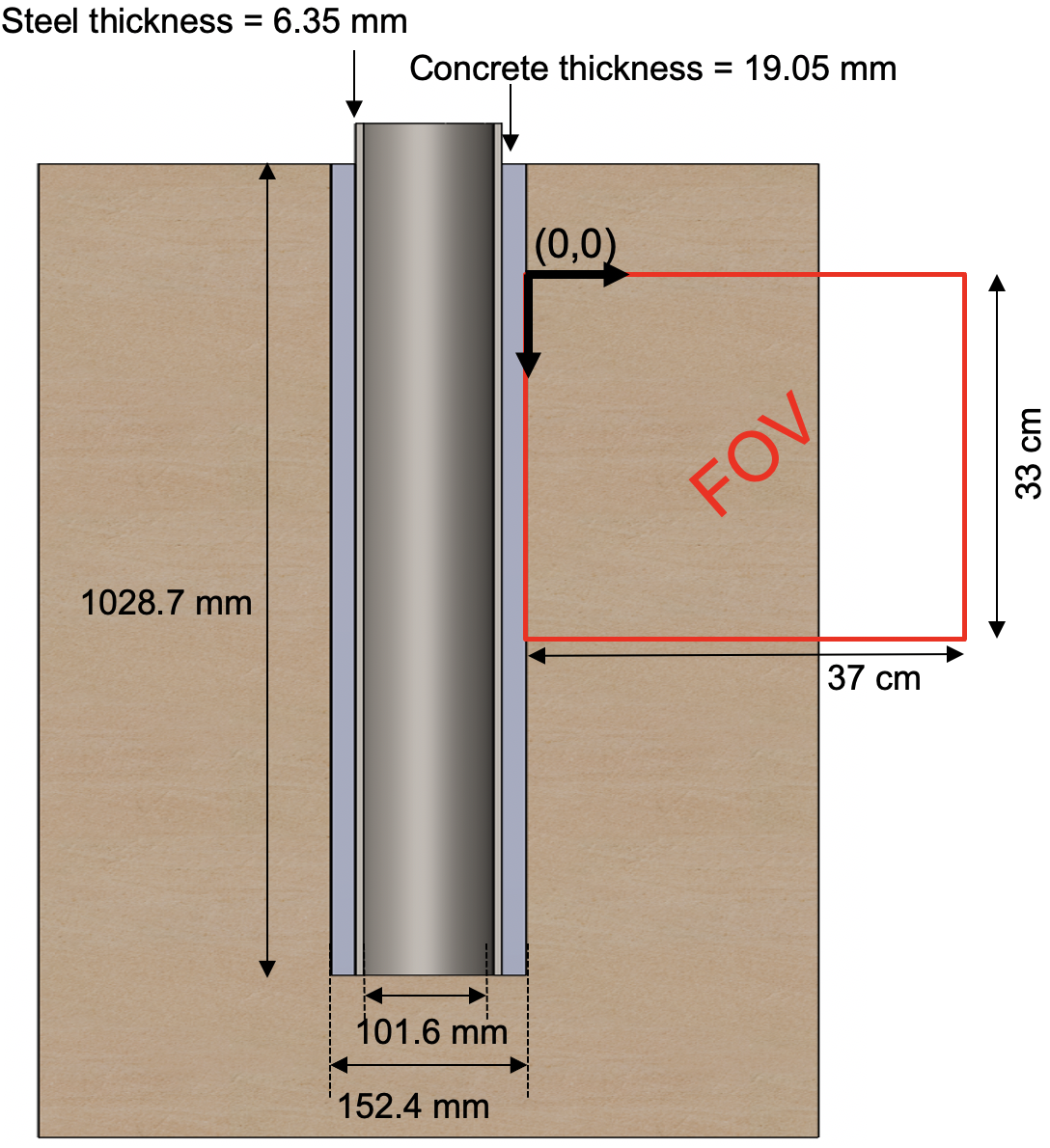}
 \tabularnewline
  \quad (a) & (b) & (c) & (d) 
 \end{tabular}
\caption{Set up for the Granite Block experiment. 
(a) Top-view photo of the granite block used in the experiment. The two left and right defects are approximately 21 cm away from the center of the borehole. 
(b) The transmitter/receiver system used in experiment. 
(c) The dimensions of the granite block and 
(d) A diagram for a cross-section. The red box indicates the area reconstructed by UMBIR.
}
\label{fig:LANL_II_obj}
\end{figure*}


Figure~\ref{fig:LANL_II_obj} illustrates the experimental set up use of the granite block data set described in this section.
The granite block has a borehole with a steel casing of 4" inner diameter and 4.5" outer diameter. 
The casing-cement interface is located at 4.5" from the borehole center, and the cement-granite interface is located at 5.25". 
The granite block has induced defects, as can be seen in Figure~\ref{fig:LANL_II_obj}(a), located at approximately 21 cm from the borehole center. 

Before imaging, the borehole was filled with water. 
Similar to the previous experiment, there is one transmitter and 15 receivers. 
The transmitter is aligned with the detectors and has a firing angle of $35^\circ$. 
The receiver/transmitter system deployed to the granite block is shown in Figure~\ref{fig:LANL_II_obj}(b).

Data was collected from 24 azimuth scans uniformly separated with an angle $15^\circ$ to cover of $360^\circ$.
Table~\ref{table:TransmitterParams_GB} lists out the transmission parameters at the three different excitation frequencies that were used for this experiment.
Table~\ref{table:UMBIR-Reconstruction-Params} gives the parameter values used in reconstructions of the granite block data, and the red box in Figure~\ref{fig:LANL_II_obj}(c) shows the area reconstructed by UMBIR.

\begin{table}[htb]
\caption{Transmit signal parameters used for the granite block experiment.}
\label{table:TransmitterParams_GB}
\centering
\setlength\tabcolsep{5pt}
\begin{tabular}{|c|c|c|} 
\hline
Excitation Frequency & Duration & Pulse Shape \\ \hline \hline
103.38 kHz  & 150  $\mu$s & Tukey \\  \hline
162.316 kHz  & 50 $\mu$s & Tukey \\  \hline
220.554 kHz  & 50 $\mu$s & Tukey \\  \hline
\end{tabular}
\end{table}

\begin{figure*}[htb]
\centering
\begin{tabular}{cccccc}
\tabularnewline
\small{103.38 kHz} & \small{162.316 kHz} & \small{220.554 kHz} & \small{MF-UMBIR}
\tabularnewline
\put(0,0)\centering{\rotatebox{90}{{\quad \quad $0^\circ$ (no defect)}}}   
\includegraphics[width=0.22\textwidth]{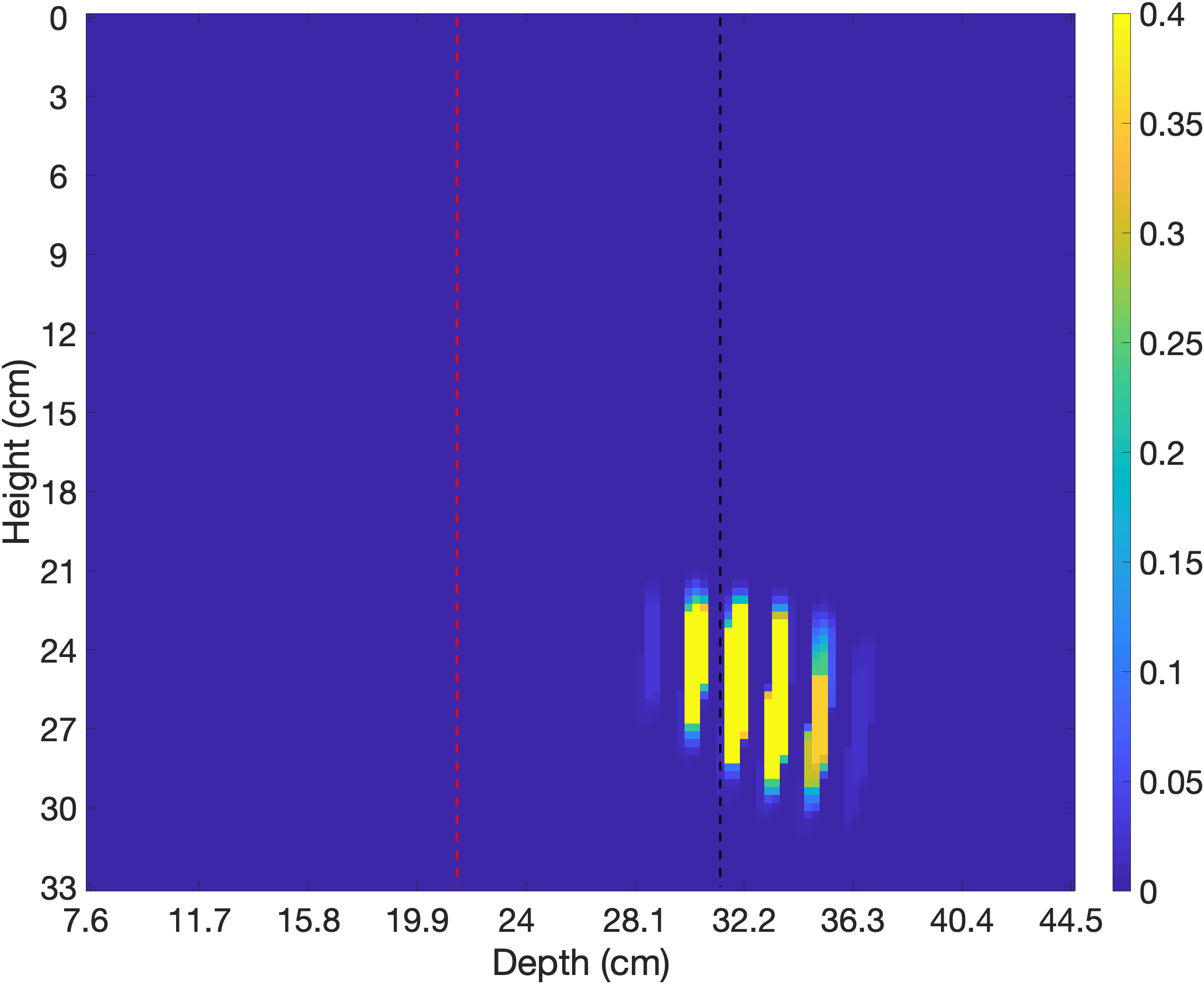} &
\includegraphics[width=0.22\textwidth]{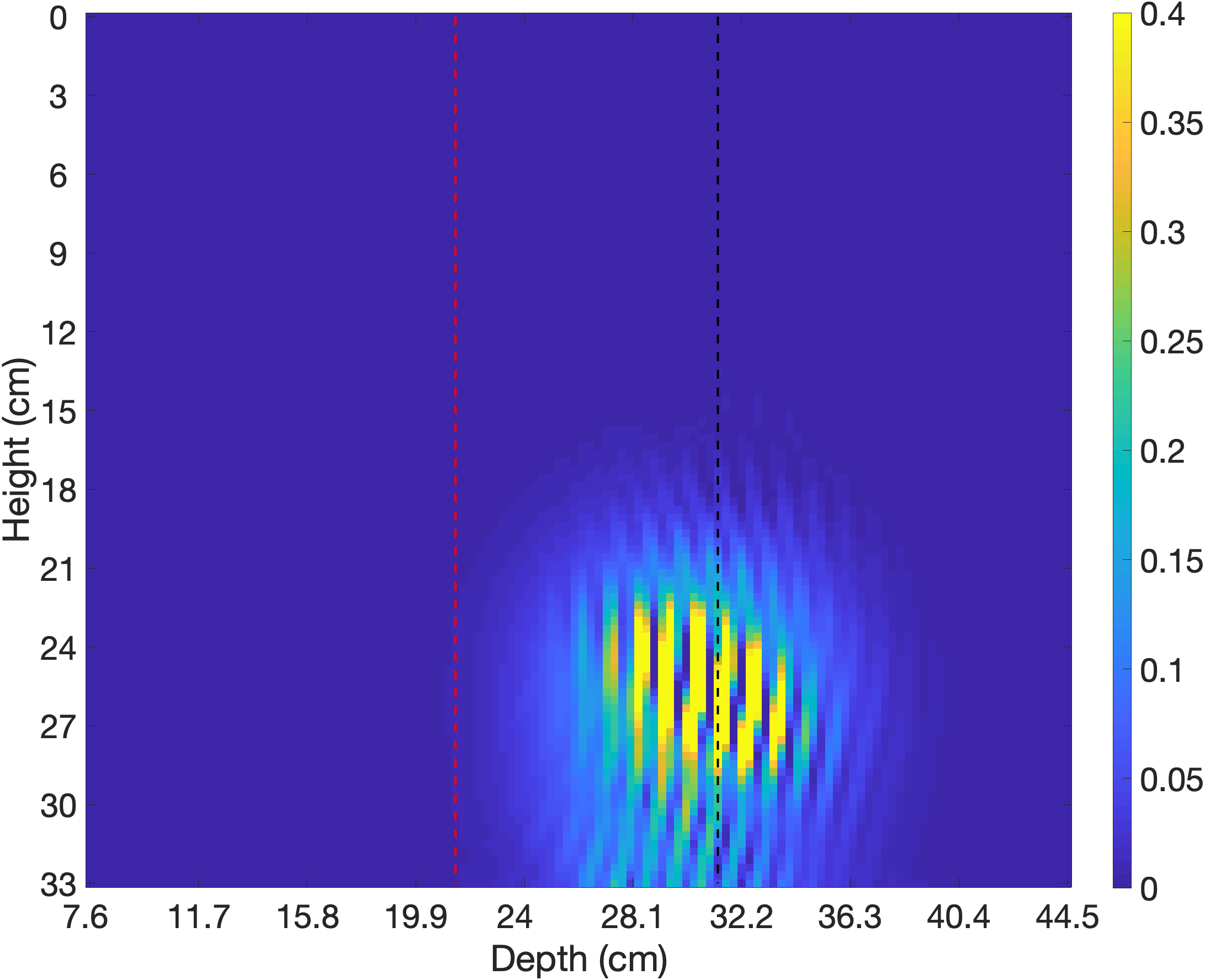} &
\includegraphics[width=0.22\textwidth]{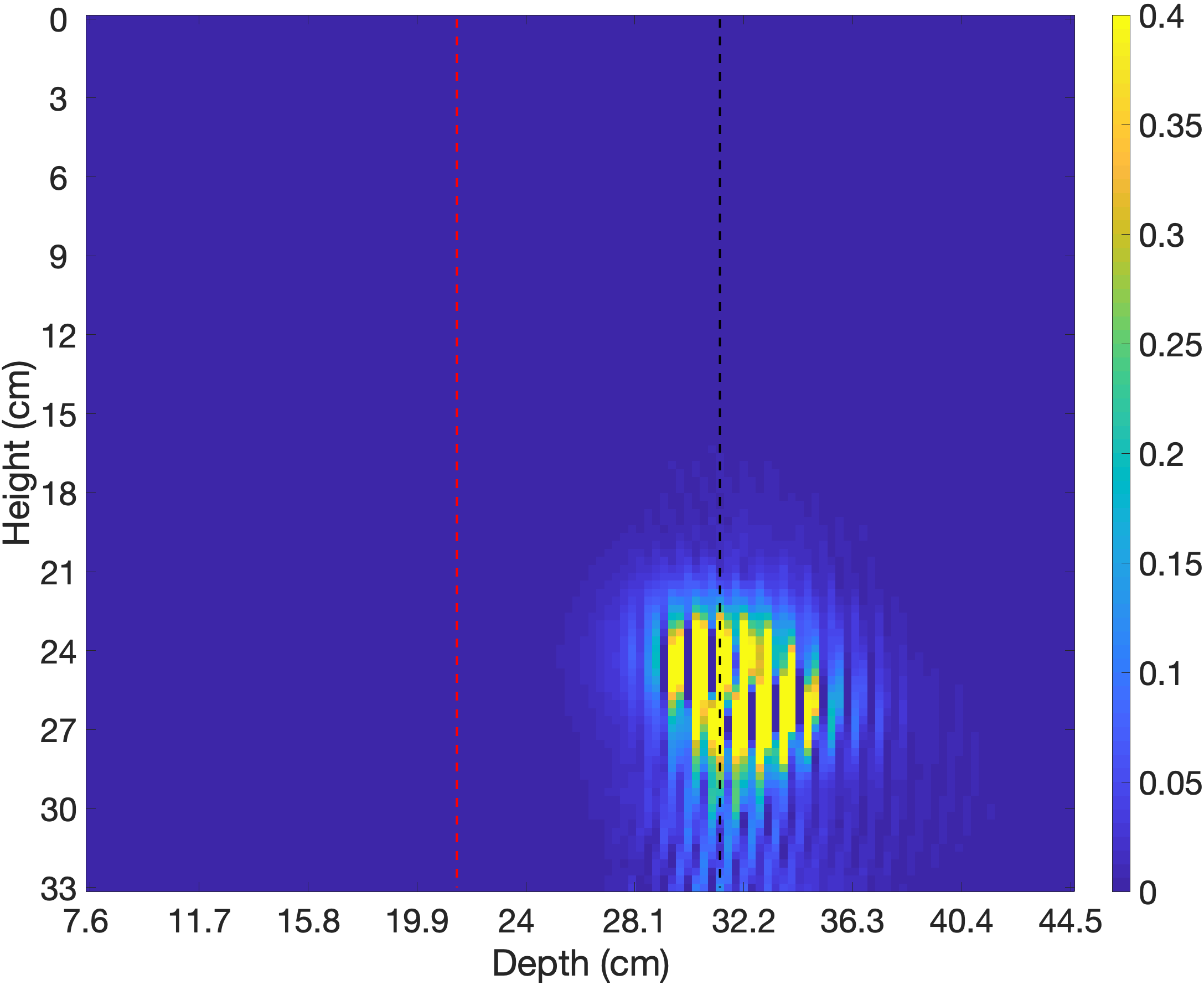} &
\includegraphics[width=0.22\textwidth]{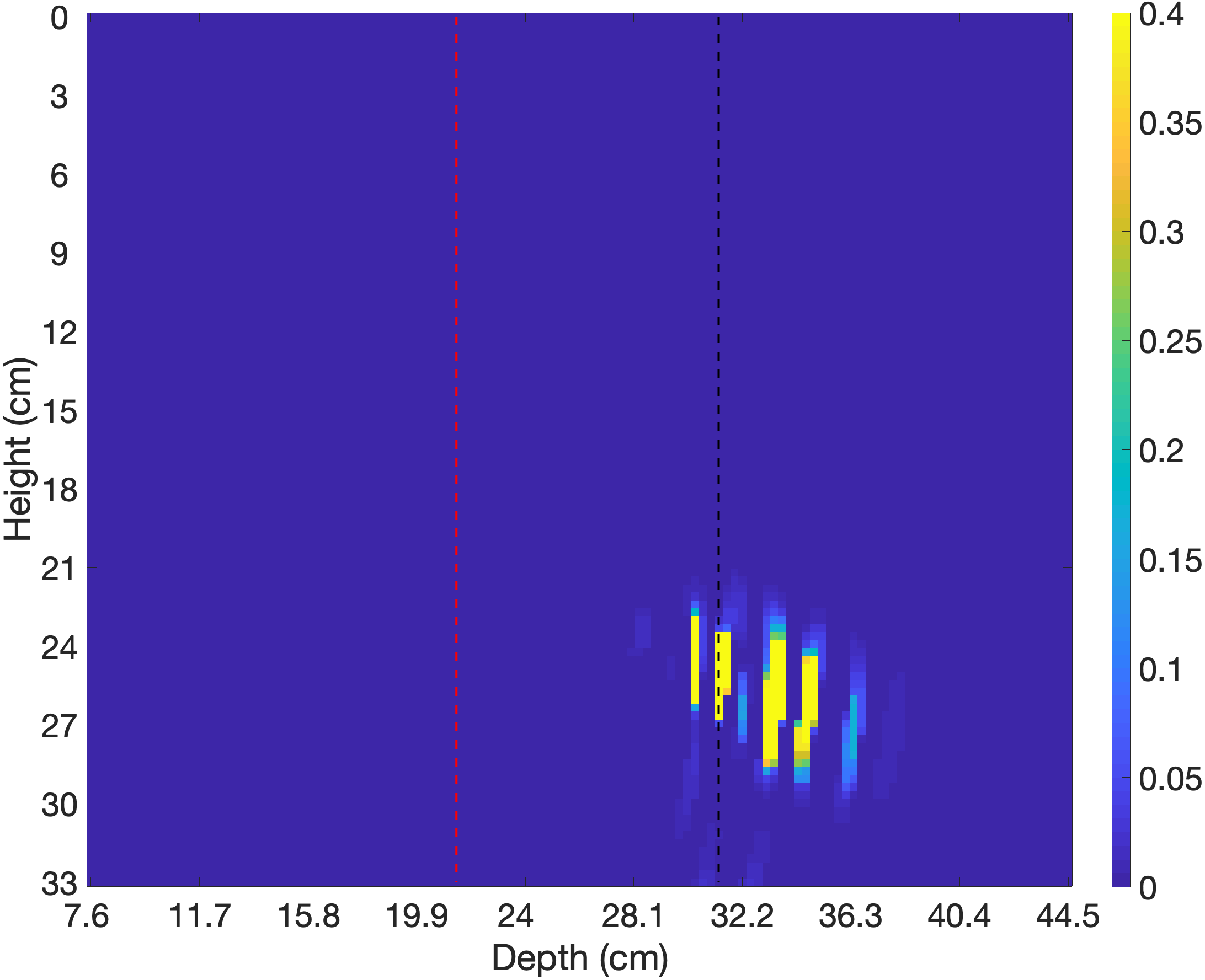} 
\tabularnewline
\put(0,0)\centering{\rotatebox{90}{{\quad \quad $90^\circ$ (defect)}}}   
\includegraphics[width=0.22\textwidth]{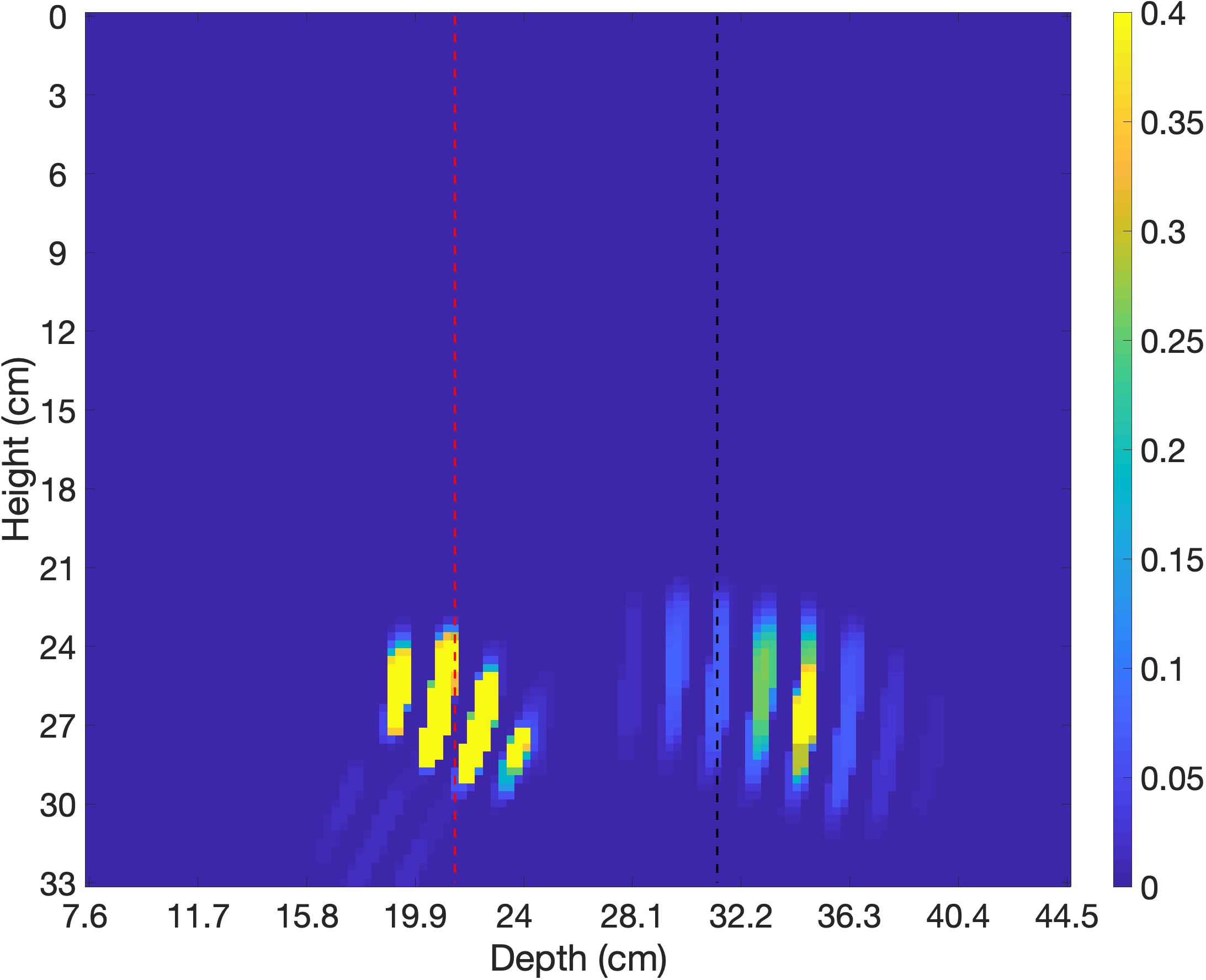} &
\includegraphics[width=0.22\textwidth]{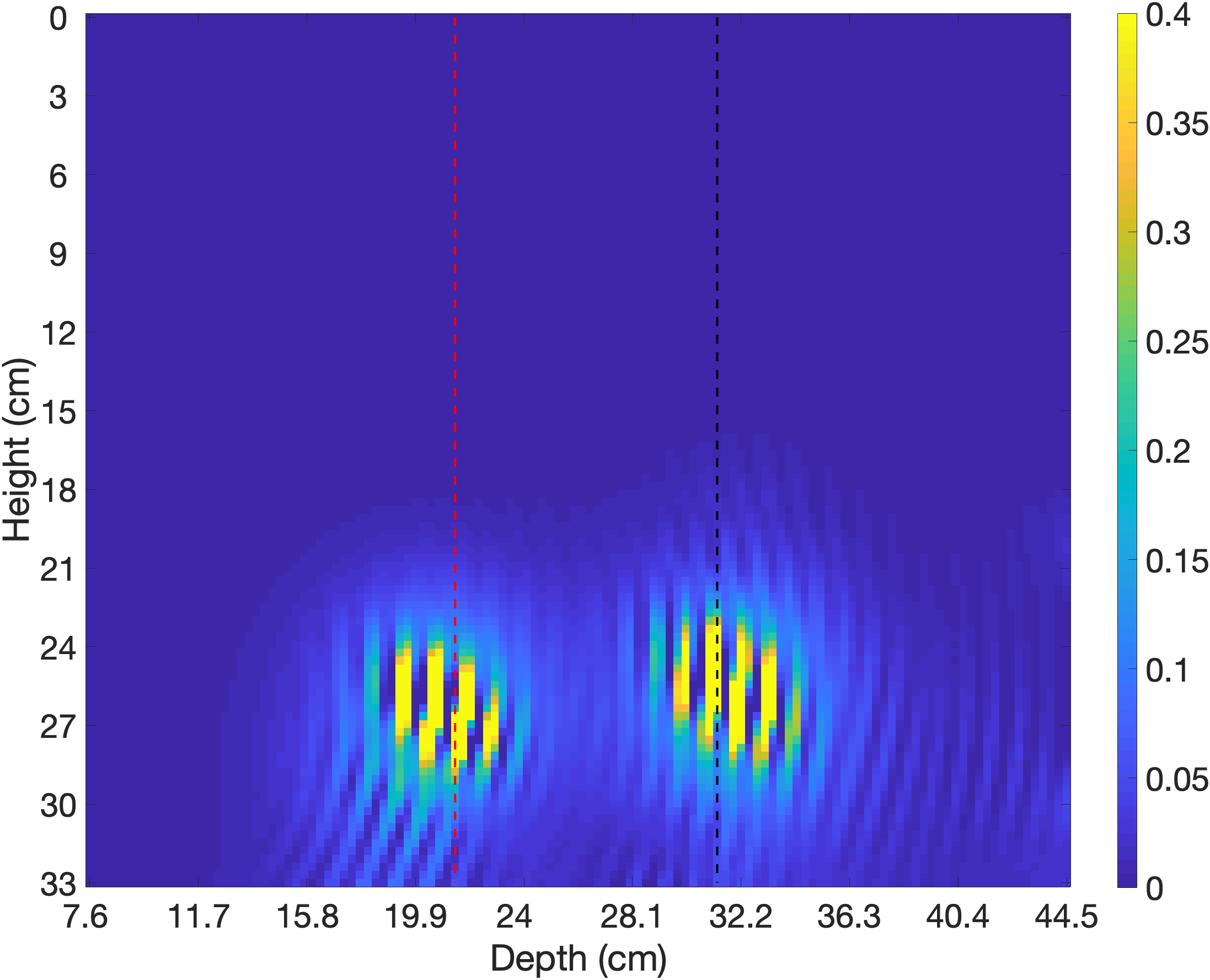} &
\includegraphics[width=0.22\textwidth]{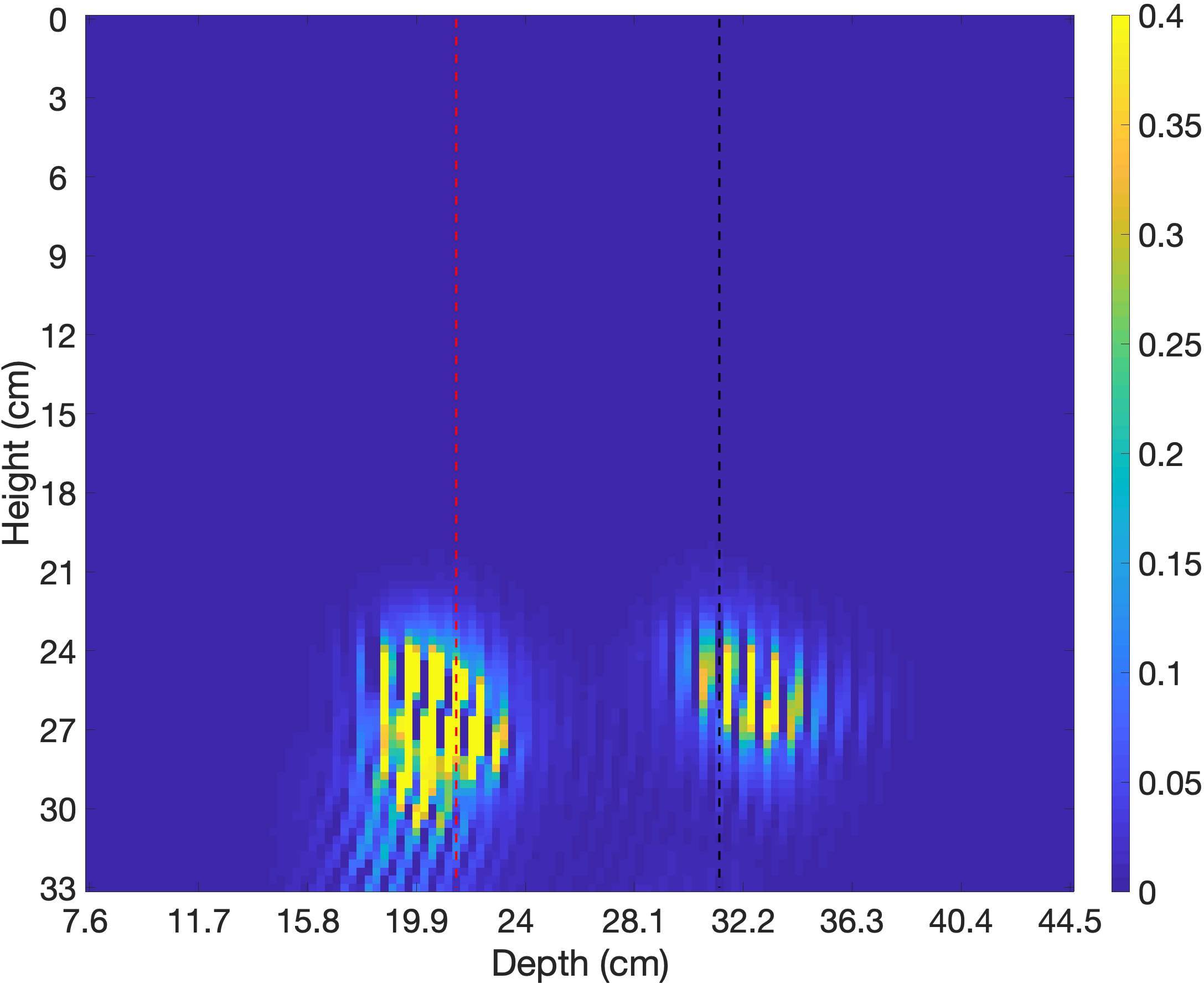} &
\includegraphics[width=0.22\textwidth]{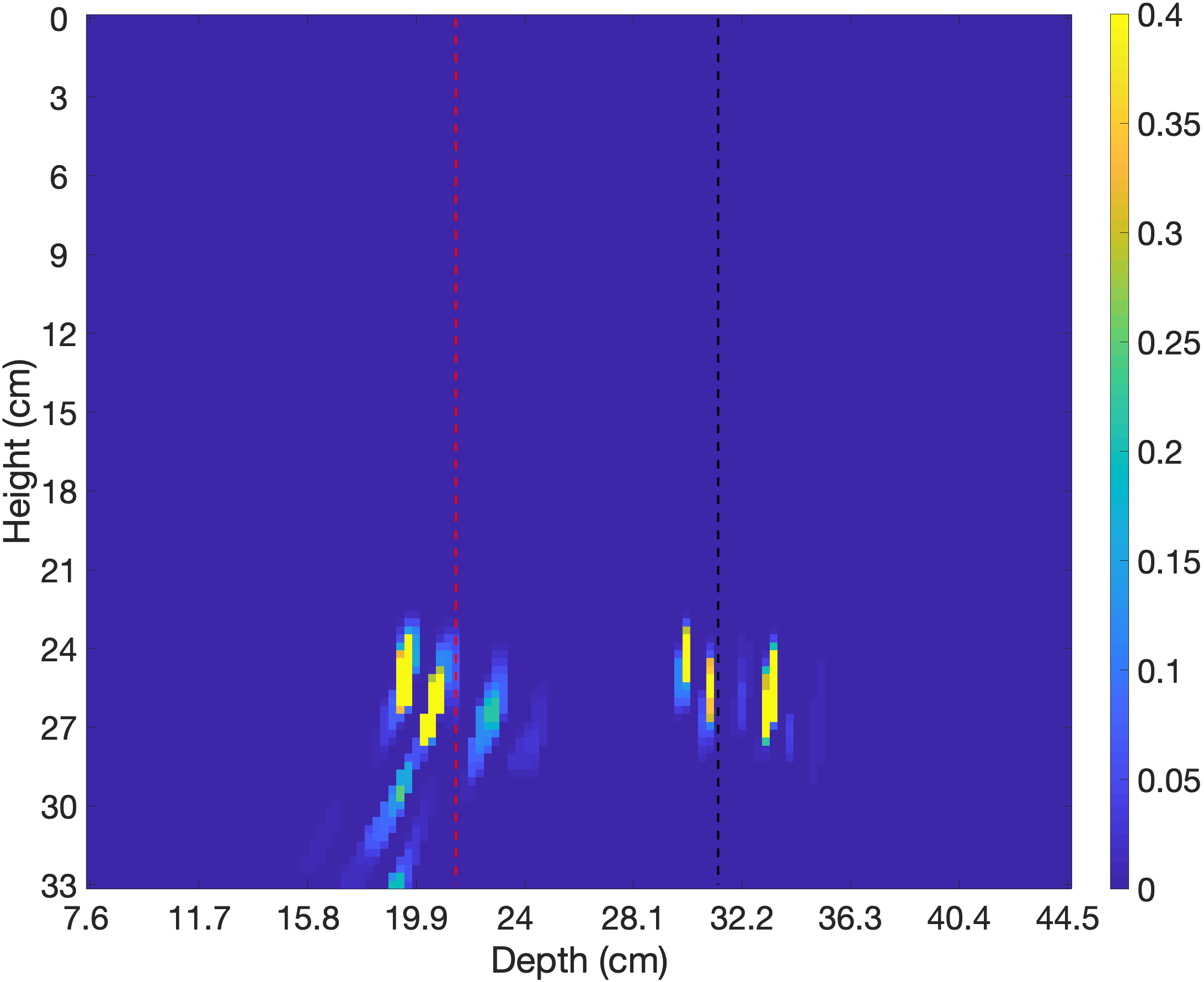} &
\tabularnewline
 \end{tabular}
\caption{UMBIR and multi-frequency UMBIR reconstruction results using experimentally measured data from the granite block (GB) experiment. 
Left to right columns correspond to the excitation frequencies 103.38 kHz, 162.316 kHz, 220.554 kHz, and the multi-frequency joint reconstructions, respectively.
The first row of reconstruction corresponds to $0^\circ$ with no defect, and the second row corresponds to $90^\circ$ with a defect.
The red doted line indicates the location of the defect.
Notice that the multi-frequency UMBIR reconstructions shows substantial improvements over single-frequency reconstructions and that it more accurately localizes the position of the defect and then wall.}
\label{fig:LANL_II_realResults}
\end{figure*}

\begin{figure}[htb]
\begin{center}
\includegraphics[width=0.5\textwidth]{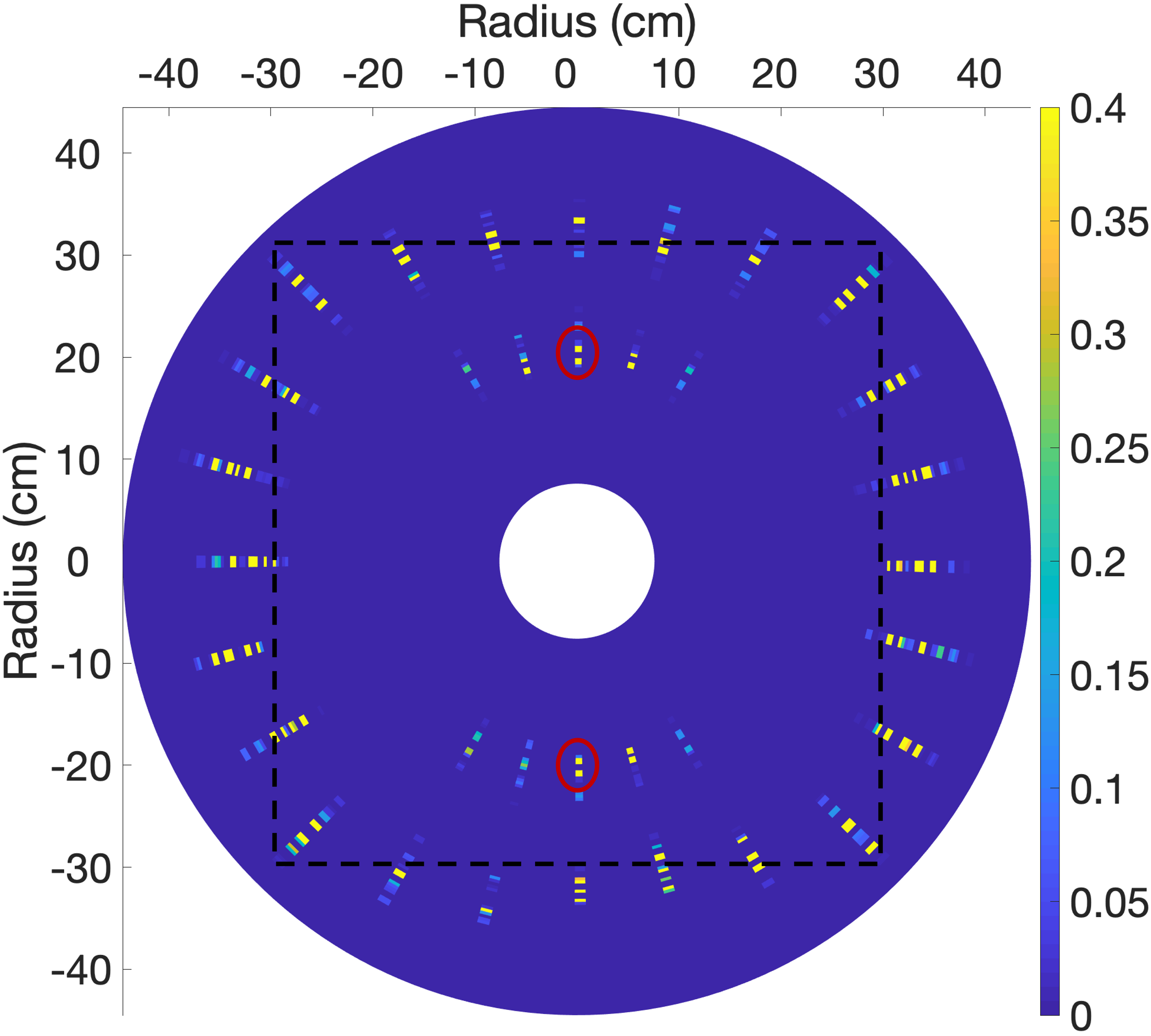}
\flushleft\caption{The multi-frequency UMBIR panoramic stitched reconstruction of the granite block (GB) at a fixed height of 25.8 cm obtained at each of 24 views in the range from $0^\circ$ to $360^\circ$ and positioned in polar coordinates. The dashed black square shows the outer edges of the GB, and the red circles indicate the two embedded defects. The defects and side walls are imaged fairly accurately, while the distance from center to each corner is underestimated, likely due to reflection effects in the corners.}
\label{fig:LANL_II_realResults_slice}
\end{center}
\end{figure}

Figure \ref{fig:LANL_II_realResults} shows cross-section reconstructions from the granite block experimental data, at rotational positions $0^\circ$ and $90^\circ$, with the defect in view at $90^\circ$ and not at $0^\circ$.
Notice that the multi-frequency UMBIR reconstruction at $0^\circ$ has a clear reflection from the sidewall around 31 cm (the black dashed line), while the reconstruction $90^\circ$ has a clear reflection from the defect around 21 cm (the red dashed line).

Figure~\ref{fig:LANL_II_realResults_slice} shows the multi-frequency UMBIR panoramic reconstruction at a fixed height of 25.8 cm in polar coordinates.  
The two red circles indicate the defect locations, and the black dashed square shows the wall location. 
Notice that the multi-frequency UMBIR reconstruction shows clear reflections from the edges of the specimen around the wall location along with the defects at $90^\circ$ and $270^\circ$. 
The distance from center to each corner is underestimated, which we speculate is due to multiple reflections near the corners.

\section{Conclusion}
\label{sec:conc}

In this paper, we proposed a multi-layer, multi-frequency collimated ultrasound model-based iterative reconstruction (UMBIR) algorithm.
To do this, we introduced a computationally efficient method for computing an accurate forward model system matrix for multi-layered structures that can typically occur in practical ultrasound imaging scenarios using collimated beam ultrasonic transducers.
We also formulated the reconstruction method using MAP reconstruction with space-varying image prior along with model of the direct arrival signal.

We tested our method on both simulated and experimentally measured data using two different scenarios corresponding to a concrete cylinder and a granite block.
Both scenarios were designed to represent the imaging of defects in a well-bore. 

In all cases, we found that multi-frequency UMBIR reconstructions had substantially better quality than single-frequency UBMIR reconstructions, and that single-frequency UMBIR reconstructions had substantially better quality than SAFT reconstructions.
Multi-frequency UMBIR reconstructions had much better localization of image defects. 
In addition, the multi-frequency UMBIR reconstructions accurately detected the location of defects and object back walls.

\section*{Acknowledgment}
A. M. Alanazi was supported by King Saud University. 
C. A. Bouman was partially supported by the Showalter Trust and by the U.S. Department of Energy. 
C. A. Bouman and G.T. Buzzard were partially supported by NSF CCF-1763896. 
S.V. and Hector Santos-Villalobos were supported by the U.S. Department of Energy staff office of the Under Secretary for Science and Energy under the Subsurface Technology and Engineering Research, Development, and Demonstration (SubTER) Crosscut program, and the office of Nuclear Energy under the Light Water Reactor Sustainability (LWRS) program.



\ifCLASSOPTIONcaptionsoff
  \newpage
\fi

\bibliographystyle{IEEEtran}
\bibliography{bare_jrnl}

\begin{thebibliography}{10}
\providecommand{\url}[1]{#1}
\csname url@samestyle\endcsname
\providecommand{\newblock}{\relax}
\providecommand{\bibinfo}[2]{#2}
\providecommand{\BIBentrySTDinterwordspacing}{\spaceskip=0pt\relax}
\providecommand{\BIBentryALTinterwordstretchfactor}{4}
\providecommand{\BIBentryALTinterwordspacing}{\spaceskip=\fontdimen2\font plus
\BIBentryALTinterwordstretchfactor\fontdimen3\font minus
  \fontdimen4\font\relax}
\providecommand{\BIBforeignlanguage}[2]{{%
\expandafter\ifx\csname l@#1\endcsname\relax
\typeout{** WARNING: IEEEtran.bst: No hyphenation pattern has been}%
\typeout{** loaded for the language `#1'. Using the pattern for}%
\typeout{** the default language instead.}%
\else
\language=\csname l@#1\endcsname
\fi
#2}}
\providecommand{\BIBdecl}{\relax}
\BIBdecl

\bibitem{chillara2017low}
V.~K. Chillara, C.~Pantea, and D.~N. Sinha, ``Low-frequency ultrasonic
  bessel-like collimated beam generation from radial modes of piezoelectric
  transducers,'' \emph{Applied Physics Letters}, vol. 110, no.~6, p. 064101,
  2017.

\bibitem{chillara2018radial}
V.~K. Chillara, C.~Pantea, and D.~Sinha, ``Radial modes of laterally stiffened
  piezoelectric disc transducers for ultrasonic collimated beam generation,''
  \emph{Wave Motion}, vol.~76, pp. 19--27, 2018.

\bibitem{chillara2019collimated}
V.~K. Chillara, E.~S. Davis, C.~Pantea, and D.~N. Sinha, ``Collimated acoustic
  beams from radial modes of piezoelectric disc transducers,'' in \emph{AIP
  Conference Proceedings}, vol. 2102, no.~1.\hskip 1em plus 0.5em minus
  0.4em\relax AIP Publishing LLC, 2019, p. 040013.

\bibitem{chillara2020ultrasonic}
V.~K. Chillara, J.~Greenhall, and C.~Pantea, ``Ultrasonic waves from radial
  mode excitation of a piezoelectric disc on the surface of an elastic solid,''
  \emph{Smart Materials and Structures}, vol.~29, no.~8, p. 085002, 2020.

\bibitem{prine1972synthetic}
D.~Prine, ``Synthetic aperture ultrasonic imaging,'' in \emph{Proceedings of
  the Engineering Applications of Holography Symposium, Los Angeles, CA, USA},
  vol. 1617, 1972.

\bibitem{stepinski2007implementation}
T.~Stepinski, ``An implementation of synthetic aperture focusing technique in
  frequency domain,'' \emph{ieee transactions on ultrasonics, ferroelectrics,
  and frequency control}, vol.~54, no.~7, pp. 1399--1408, 2007.

\bibitem{hoegh2015extended}
K.~Hoegh and L.~Khazanovich, ``Extended synthetic aperture focusing technique
  for ultrasonic imaging of concrete,'' \emph{NDT \& E International}, vol.~74,
  pp. 33--42, 2015.

\bibitem{skjelvareid2011synthetic}
M.~H. Skjelvareid, T.~Olofsson, Y.~Birkelund, and Y.~Larsen, ``Synthetic
  aperture focusing of ultrasonic data from multilayered media using an omega-k
  algorithm,'' \emph{IEEE transactions on ultrasonics, ferroelectrics, and
  frequency control}, vol.~58, no.~5, pp. 1037--1048, 2011.

\bibitem{lin2018ultrasonic}
S.~Lin, S.~Shams, H.~Choi, and H.~Azari, ``Ultrasonic imaging of multi-layer
  concrete structures,'' \emph{NDT \& E International}, vol.~98, pp. 101--109,
  2018.

\bibitem{cerveny2005seismic}
V.~Cerveny, \emph{Seismic ray theory}.\hskip 1em plus 0.5em minus 0.4em\relax
  Cambridge university press, 2005.

\bibitem{margrave2019numerical}
G.~F. Margrave and M.~P. Lamoureux, \emph{Numerical methods of exploration
  seismology: With algorithms in MATLAB{\textregistered}}.\hskip 1em plus 0.5em
  minus 0.4em\relax Cambridge University Press, 2019.

\bibitem{shlivinski2007defect}
A.~Shlivinski and K.~Langenberg, ``Defect imaging with elastic waves in
  inhomogeneous--anisotropic materials with composite geometries,''
  \emph{Ultrasonics}, vol.~46, no.~1, pp. 89--104, 2007.

\bibitem{gazdag1978wave}
J.~Gazdag, ``Wave equation migration with the phase-shift method,''
  \emph{Geophysics}, vol.~43, no.~7, pp. 1342--1351, 1978.

\bibitem{skjelvareid2010ultrasound}
M.~H. Skjelvareid and Y.~Birkelund, ``Ultrasound imaging using multilayer
  synthetic aperture focusing,'' in \emph{Pressure Vessels and Piping
  Conference}, vol. 49248, 2010, pp. 379--387.

\bibitem{ozkan2017inverse}
E.~Ozkan, V.~Vishnevsky, and O.~Goksel, ``Inverse problem of ultrasound
  beamforming with sparsity constraints and regularization,'' \emph{IEEE
  transactions on ultrasonics, ferroelectrics, and frequency control}, vol.~65,
  no.~3, pp. 356--365, 2017.

\bibitem{tuysuzoglu2012sparsity}
A.~Tuysuzoglu, J.~M. Kracht, R.~O. Cleveland, M.~C{\c{}}~etin, and W.~C. Karl,
  ``Sparsity driven ultrasound imaging,'' \emph{The Journal of the Acoustical
  Society of America}, vol. 131, no.~2, pp. 1271--1281, 2012.

\bibitem{liu2016least}
X.~Liu, Y.~Liu, X.~Huang, and P.~Li, ``Least-squares reverse-time migration
  with cost-effective computation and memory storage,'' \emph{Journal of
  Applied Geophysics}, vol. 129, pp. 200--208, 2016.

\bibitem{zhang2015stable}
Y.~Zhang, L.~Duan, and Y.~Xie, ``A stable and practical implementation of
  least-squares reverse time migration,'' \emph{Geophysics}, vol.~80, no.~1,
  pp. V23--V31, 2015.

\bibitem{chen2018full}
Y.~Chen, K.~Gao, E.~S. Davis, D.~N. Sinha, C.~Pantea, and L.~Huang,
  ``Full-waveform inversion and least-squares reverse-time migration imaging of
  collimated ultrasonic-beam data for high-resolution wellbore integrity
  monitoring,'' \emph{Applied Physics Letters}, vol. 113, no.~7, p. 071903,
  2018.

\bibitem{zeng2017guide}
C.~Zeng, S.~Dong, and B.~Wang, ``A guide to least-squares reverse time
  migration for subsalt imaging: Challenges and solutions,''
  \emph{Interpretation}, vol.~5, no.~3, pp. SN1--SN11, 2017.

\bibitem{wu2015model}
H.~Wu, J.~Chen, S.~Wu, H.~Jin, and K.~Yang, ``A model-based regularized inverse
  method for ultrasonic b-scan image reconstruction,'' \emph{Measurement
  Science and Technology}, vol.~26, no.~10, p. 105401, 2015.

\bibitem{almansouri2018model}
H.~Almansouri, S.~Venkatakrishnan, C.~Bouman, and H.~Santos-Villalobos,
  ``Model-based iterative reconstruction for one-sided ultrasonic
  nondestructive evaluation,'' \emph{IEEE Transactions on Computational
  Imaging}, vol.~5, no.~1, pp. 150--164, 2018.

\bibitem{almansouri2018anisotropic}
H.~Almansouri, S.~Venkatakrishnan, D.~Clayton, Y.~Polsky, C.~Bouman, and
  H.~Santos-Villalobos, ``Anisotropic modeling and joint-map stitching for
  improved ultrasound model-based iterative reconstruction of large and thick
  specimens,'' in \emph{AIP Conference Proceedings}, vol. 1949 (1).\hskip 1em
  plus 0.5em minus 0.4em\relax AIP Publishing LLC, 2018, p. 030002.

\bibitem{weston2012time}
M.~Weston, P.~Mudge, C.~Davis, and A.~Peyton, ``Time efficient auto-focussing
  algorithms for ultrasonic inspection of dual-layered media using full matrix
  capture,'' \emph{Ndt \& E International}, vol.~47, pp. 43--50, 2012.

\bibitem{moser1991shortest}
T.~Moser, ``Shortest path calculation of seismic rays,'' \emph{Geophysics},
  vol.~56, no.~1, pp. 59--67, 1991.

\bibitem{sethian1996fast}
J.~A. Sethian, ``A fast marching level set method for monotonically advancing
  fronts.'' \emph{Proceedings of the National Academy of Sciences}, vol.~93,
  no.~4, pp. 1591--1595, 1996.

\bibitem{brath2017phased}
A.~J. Brath and F.~Simonetti, ``Phased array imaging of complex-geometry
  composite components,'' \emph{IEEE Transactions on Ultrasonics,
  Ferroelectrics, and Frequency Control}, vol.~64, no.~10, pp. 1573--1582,
  2017.

\bibitem{alanazi2022model}
A.~Alanazi, S.~Venkatakrishnan, H.~Santos-Villalobos, G.~Buzzard, and
  C.~Bouman, ``Model-based reconstruction for collimated beam ultrasound
  systems,'' in \emph{ICASSP 2022-2022 IEEE International Conference on
  Acoustics, Speech and Signal Processing (ICASSP)}.\hskip 1em plus 0.5em minus
  0.4em\relax IEEE, 2022, pp. 1601--1605.

\bibitem{bouman2022foundations}
C.~A. Bouman, \emph{Foundations of Computational Imaging: A Model-Based
  Approach}.\hskip 1em plus 0.5em minus 0.4em\relax SIAM, 2022, vol. 180.

\bibitem{thibault2007three}
J.-B. Thibault, K.~D. Sauer, C.~A. Bouman, and J.~Hsieh, ``A three-dimensional
  statistical approach to improved image quality for multislice helical ct,''
  \emph{Medical physics}, vol.~34, no.~11, pp. 4526--4544, 2007.

\bibitem{treeby2010k}
B.~E. Treeby and B.~T. Cox, ``k-wave: Matlab toolbox for the simulation and
  reconstruction of photoacoustic wave fields,'' \emph{Journal of biomedical
  optics}, vol.~15, no.~2, p. 021314, 2010.

\bibitem{pantea2019collimated}
C.~Pantea, ``Collimated beams for cement evaluation,'' Los Alamos National
  Lab.(LANL), Los Alamos, NM (United States), Tech. Rep., 2019.

\end{thebibliography}

\end{document}